\documentclass[fleqn,nofootinbib,showkeys,pra]{iopart}
\usepackage{setspace}
\usepackage{amssymb}
\usepackage{xcolor}
\usepackage{color}
\usepackage{graphicx}

\usepackage[normalem]{ulem}
\newcommand{\R}{{\mathbb{R}}}

\newcommand{\N}{{\mathbb{N}}}

\begin{document}
\title{Bound states and point interactions of the one-dimensional pseudospin-one Hamiltonian}
\author{A.~V. Zolotaryuk, Y. Zolotaryuk\footnote{Author to whom any correspondence should be addressed} and V.~P. Gusynin}
\address{Bogolyubov Institute for Theoretical Physics, National Academy of
Sciences of Ukraine, vul. Metrologichna 14b, Kyiv 03143, Ukraine}
\ead{yzolo@bitp.kiev.ua}

\date{\today}

\begin{abstract}
The spectrum of a  one-dimensional pseudospin-one Hamiltonian with a three-component potential is studied for two configurations: (i) all the potential components are
constants over the whole coordinate space and (ii) the profile of some components
is of a rectangular form.
In case (i), it is illustrated how the  structure of three (lower, middle and upper)
bands  depends  on the configuration of potential strengths including the appearance of
flat bands at some special values of these strengths. In case (ii), the set of two
 equations for finding bound states is derived.  The spectrum of bound-state
 energies is shown to depend crucially
 on the configuration of potential strengths. Each of these configurations is
 specified by a  single strength parameter $V$. The bound-state energies are
 calculated as functions
 of the strength $V$ and a one-point approach is developed realizing correspondent point
 interactions.  For different potential configurations, the energy dependence
 on the strength $V$ is described in detail, including its one-point approximation.
From a whole variety of bound-state spectra, four characteristic types are singled out.

\end{abstract}


\submitto{\JPA}

\section{Introduction}

Experimental discovery of graphene attracted attention to condensed matter systems with spectrum of quasiparticles similar to the relativistic
one. It is well known that quasiparticle excitations in graphene are described at low energies by the massless Dirac equation in two space dimensions.
Moreover, it was shown \cite{Bradlyn} that more complicated fermionic quasiparticles
could be realized in crystals with special space groups with
no analogues in particle physics, where the Poincar\'{e} symmetry provides strong restrictions allowing only three types: Dirac, Weyl and Majorana
(not discovered yet) particles with spin 1/2. In condensed matter systems, besides fermions with pseudospin 1/2, other fermions with a higher pseudospin
can appear in two- and three-dimensional solids. In particular, special attention is paid
to fermionic excitations with pseudospin one, whose
Hamiltonian is given by the scalar product of momentum and the spin-1 matrices \cite{Bercioux2009,Raoux2014,laf}.

Many aspects of pseudospin-1 Hamiltonians, such as the energy spectrum having a flat band along with two dispersive bands which are linear in momentum
as in graphene, are fascinating. The dice model is an example of such a system, which hosts pseudospin-1 fermions with a completely flat band at zero
energy \cite{Bercioux2009,Raoux2014}.  The quenching of the kinetic energy in flat bands strongly enhances the role of electron-electron and other
interactions and may lead to the realization of many very interesting correlated states such as ferromagnetism \cite{Tasaki1998}, superconductivity
in twisted bilayer graphene \cite{Cao2018} and plethora of other quantum phases \cite{Andrei2020}.

Currently, a whole body of literature has been accumulated, which is devoted to  the investigation of physical quantities in the presence of flat bands
in two-dimensional systems such as orbital susceptibility \cite{Raoux2014}, optical conductivity \cite{Illes2015,Kovacs2017,Iurov2020}, magnetotransport \cite{Biswas2016,Islam2017}, RKKY \cite{Oriekhov2020,Roslyak2021} and Coulomb \cite{Oriekhov2019,Pottelberge2020} interactions. However, one-dimensional
pseudospin-1 systems have been much less studied. Here, one should note the recent works by Zhang with coauthors \cite{Zhang2022,Zhang2022JPB,Zhang2022PS},
where the bound state problem in a one-dimensional pseudospin-1 Dirac Hamiltonian with a flat band  was investigated in the presence of delta- and square
well potentials.  In particular, the existence of infinite series of bound states near the flat band appears to be of great interest 
\cite{Zhang2022JPB}. Very recently, the transport properties and snake states of pseudospin-1 Dirac-like electrons have been analyzed  by Jakubsk\'{y} 
and  Zelaya \cite{Jakubsky2023_1,Jakubsky2023_2} in Lieb lattice under barrier- and well-like electrostatic interactions. 

In the present work we consider a one-dimensional spin-1 Hamiltonian $H=H_0 +V(x)$ with its free-particle part
\begin{equation}
\!\!\!\!\!\!\!\!
\hspace{-20pt}H_0 = - {\rm i} S_y {d \over dx} +m S_z\,,~~S_y = {1 \over \sqrt{2}}
\left(\begin{array}{lll}
0~-{\rm i}~~~0 \\ {\rm i}~~~~0~-{\rm i}\\ 0~~~~{\rm i}~~~~0
 \end{array} \right),
~~S_z =\left(  \begin{array}{lll} 1~~~0~~~~0  \\ 0~~~0~~~~0 \\ 0~~~0~ \!-1
\end{array} \right)
\label{free_Hamiltonian}
\end{equation}
and a potential
\begin{equation}
 V(x)= \left( \begin{array}{ccc} V_{11}(x)~~~~~~0~~~~~~~~~~\,0 \\
~~~0~~~~~~~V_{22}(x)~~~~~~\,0 \\ ~~~~~~~\,0~~~~~~~~~~0~~~~~~~~V_{33}(x)
              \end{array} \right).
 \label{potential}
 \end{equation}
We use the matrix $S_y$ instead of $S_x$ in \cite{Zhang2022JPB} in order to have real coefficients in the equations as
  the Dirac equation in the Majorana representation.

Let $\psi(x) = {\rm col}\!\left( \psi_1(x),\, \psi_2(x),\,\psi_3(x)\right)$ be
a three-component wave function.
Then the Schr\"{o}dinger equation $\left[H_0 +V(x)\right]\psi(x)= E\psi(x)$ with
energy $E$ is represented in the component form as the system of three equations:
\begin{equation}
 \begin{array}{lll} \smallskip
 -\, \psi'_2(x)/\sqrt{2} +\left[m +V_{11}(x)\right]
 \psi_1(x)\! &=& \!E \psi_1(x),\\ \smallskip
  \left[\psi'_1(x) -\psi'_3(x)\right]/\sqrt{2} + V_{22}(x)
 \psi_2(x)\!&=&\! E\psi_2(x), \\
  \psi'_2(x)/\sqrt{2} - \left[m - V_{33}(x)\right]
 \psi_3(x)\!& =&\! E \psi_3(x) \,,\end{array}
 \label{basic_eqs}
\end{equation}
where the prime stands for the differentiation over $x$. Notice that adding the first and third equations we get an algebraic relation
between the functions $\psi_1$ and $\psi_3$. In fact, we have two differential equations and  one algebraic constraint. This is due to the fact
that the matrix $S_y$ is singular, ${\rm det}S_y=0$, and its rank equals two.
Thus the system (\ref{basic_eqs}) cannot be transformed to the canonical
form $\dot\psi_i=M_{ij}\psi_j$ for the system of differential equations.

The free-particle spectrum of equations (\ref{basic_eqs}),
where $V_{11}(x)=V_{22}(x)=V_{33}(x) \equiv 0$, consists of the three bands:
\begin{equation}
\!\!\!\!\!\!\!\!\!\!\!\!\!\!\!\!\!\!\!\!\!\!\!\!\!\!\!\!
 E=0~(\mbox{flat band}),~E= \pm \sqrt{k^2 +m^2}~(\mbox{upper and lower
 dispersion bands}).
 \label{free_bands}
\end{equation}
The gap in this spectrum consists of the two intervals $-m < E < 0$ and $0 <E<m$ where possible bound states can exist
in the presence of a potential term. The spectrum of the Hamiltonian $H_0$ is
particle-hole symmetric with the isolated flat band
at zero energy, which is a consequence of the existence of a matrix $C$,
\begin{eqnarray}
C=\left(\begin{array}{ccc}0&0&1\\0&1&0\\ 1&0&0\end{array}\right),
\end{eqnarray}
that anti-commutes with $H_0$. It is interesting that, for another type of the mass term
$m \,{\rm diag}(1,-1,1)$,
the flat band with the energy $E=m$ exists and touches either the upper ($m > 0$) or the lower ($m < 0$) dispersive energy band,
thus violating the particle-hole symmetry (similar to the
two-dimensional $\alpha-{\cal T}_3$ model \cite{Piechon2015,Oriekhov2019}).

While in the non-relativistic case, in the presence of an external constant potential,
the free-particle spectrum is simply shifted
accordingly, the spectrum of system (\ref{basic_eqs}), in a similar situation where the strength components $(V_{11}$, $V_{22}$ and
$V_{33})$ are constant over the whole $x$-axis, depends on the configuration of these components in a non-trivial way. Therefore it
is of interest to examine the spectrum structure of a pseudospin-1 Hamiltonian depending
on all the vectors
${\rm col}(V_{11},\, V_{22}, \,V_{33})$ which forms a three-dimensional space $\R^3$.
The further task is to single out explicitly in this space the sets of the existence
of flat bands.

For realizing bound states  of the pseudospin-one Hamiltonian,  the components
$V_{11}(x)$,  $V_{22}(x)$ and $V_{33}(x)$, defined as functions on the whole $x$-axis,
must decay to zero at $|x| \to \infty$.  Then the bound states (if any)
are expected to appear within the gap $-m < E< m$. Having the explicit solution
of equations (\ref{basic_eqs}) with constant strength components, it is reasonable
to choose the components of the potential $V(x)$ in the form of rectangles
(barriers or wells). In simple terms, such rectangular potentials describe a
heterostructure composed of parallel plane layers. The particle
motion in these systems is confined only along the $x$-axis, being free in
(perpendicular) planes. In this case, for some special configurations of the strengths
$V_{11}$,  $V_{22}$ and $V_{33}$, it is possible to examine the bound-state spectrum
in an explicit form, exhibiting a number of interesting and intriguing features.

Because of the rapid progress in fabricating nanoscale quantum devices,
the investigation of extremely thin layers described by sharply localized potentials
is of particular interest nowadays. In this regard, the so-called zero-range or
point interaction models,  which are widely used in various applications
to quantum physics \cite{Demkov1975,Demkov1988,Albeverio2005,Albeverio1999}, should also be elaborated for Dirac-like systems. In general, a point interaction, being a
singular object, is determined by the two-sided boundary conditions on a wave
function, which are given at the point of singularity (say, e.g., $x=\pm 0$).
In the case of a heterostructure consisting of a finite number of parallel layers,
it is quite useful to apply the transfer matrix approach as  a starting point to
implement such a modeling. Knowing the matrix that connects the values of a wave function and its derivative (in
the non-relativistic case) or the components of a spinor (in the relativistic
case) given at the boundaries of each mono-layer, the full transfer matrix of the
system can easily be calculated as the product of all the mono-layer matrices.
The further step is to shrink the thickness of the full multi-layered system to zero.
For example, in this way, the exactly solvable model has been constructed
for the non-relativistic Schr\"{o}dinger equation with a delta derivative potential
$\delta'(x)$ \cite{ZZ2011,ZZ2015}. In other studies, performed for instance in
\cite{Gusynin2022},  the squeezing limit This squeezed ...may be applied separately
to each layer, fixing the distances between the layers. Here, using the transfer
matrix method, the bound states of a one-dimensional Dirac equation with multiple
delta potentials have been studied. Based on this method, for a similar equation
 written in a more general form,
 the continuity between the states of perfect transmission and bound states has been established in \cite{Ibarra2023}.

Finally, it should be emphasized that the squeezed connection matrices, which
 define the corresponding point interactions, depend on the shape of the functions
used for the squeezing limit realization. This non-uniqueness problem refers to
both the non-relativistic Schr\"{o}dinger equation with a $\delta'$-like
potential \cite{zpi,Golovaty2009,Golovaty2013} and the relativistic
Dirac equation with a $\delta$-potential \cite{Gusynin2022}.
In this regard, the piecewise representation of the potential profile of layers
seems to be motivated from a physical point of view because of satisfying the
principle of strength additivity \cite{Gusynin2022}.

\section{Three-band structure of the energy spectrum
for the Hamiltonian with constant potentials }

Consider the system for which $V_{jj}(x) \equiv V_{jj}=$ const., $V_{jj} \in \R$,
$j=1,2,3$, and rewrite equations (\ref{basic_eqs}) in the form
\begin{equation}
 \begin{array}{ccc} \smallskip
 - \,\psi'_2(x) &=& \sqrt{2}\, (E-V_1) \psi_1(x),\\ \smallskip
 \psi'_1(x) -\psi'_3(x) &=&  \sqrt{2} \,(E-V_2) \psi_2(x),  \\ \smallskip
 \psi'_2(x) & =& \sqrt{2}\, (E-V_3) \psi_3(x), \end{array}
 \label{basic_eqs_const}
\end{equation}
where
\begin{equation}
V_1 := V_{11}+m,~~~V_2 \equiv V_{22}\,,~~~V_3 := V_{33}-m
\label{V_j}
\end{equation}
are renormalized potentials strengths (or intensities).

Assume that $\psi(k;x) = {\rm col}\!\left(A_1,\,A_2,\,A_3\right)\exp(\pm{\rm i}kx)$ with unknowns $A_j$'s  and a wave number $k$ being real or imaginary. Inserting this
representation  into equations (\ref{basic_eqs_const}), we get a system
of three linear equations.
 Calculating next the determinant  of this system, we arrive at the  equation
\begin{equation}
\!\!\!\!\!\!\!\!\!\!\!\!\!\!\!\!\!\!\!\!\!
 F(E)= G(E)k^2,~~~k^2 \in \R\setminus \{0\},\;\;G(E) =  E - V_a\,,~~~
 V_a := {1 \over 2}  \left( V_1  +V_{3}\right),
 \label{law}
\end{equation}
where the function $F(E)$ has  a cubic factorized form:
\begin{equation}
F(E) = (E-V_1)(E-V_2)(E-V_3)
\label{F}
\end{equation}
Cubic equation (\ref{law}) with (\ref{F}) determines the dispersion laws that describe
the relation
between the energy $E$ and the wave number $k$. Explicit solutions can be written using
the well-known formulas for the roots of this cubic equation,
but we prefer a qualitative analysis of the equation, which we will consider in the next section.

A general solution of equations (\ref{basic_eqs_const}) can be found if we
express the constants $A_1$ and $A_3$ through $A_2$,
using also equation (\ref{law}). As a result, it can be represented as the sum of
linearly independent solutions:
\begin{equation}
 \psi(k;x) = B_1 \, {\rm col}\left( -\sigma_1,\,1 ,\, \sigma_3\right)
 {\rm e}^{{\rm i}k x}+ B_2\, {\rm col}\left( \sigma_1, \,1,\, -\sigma_3 \right)
 {\rm e}^{-{\rm i}k x},
 \label{gen_sol}
\end{equation}
where the constants $B_1$ and $B_2$ are arbitrary,
\begin{equation}
 \sigma_j := {{\rm i}k \over \sqrt{2}\,(E-V_j)}\,, \quad j=1,3,
 \label{sigma_j}
\end{equation}
and the wave number $k$, being  real or imaginary, is related to the energy $E$
through the formula
\begin{equation}
 k = \sqrt{(E-V_1)(E-V_2)(E-V_3) \over E- V_a}  \,.
\label{k}
\end{equation}

\subsection{Structure of dispersion and flat bands}

\begin{figure}[htb]
\hspace{-2pt}\includegraphics[scale=0.54]{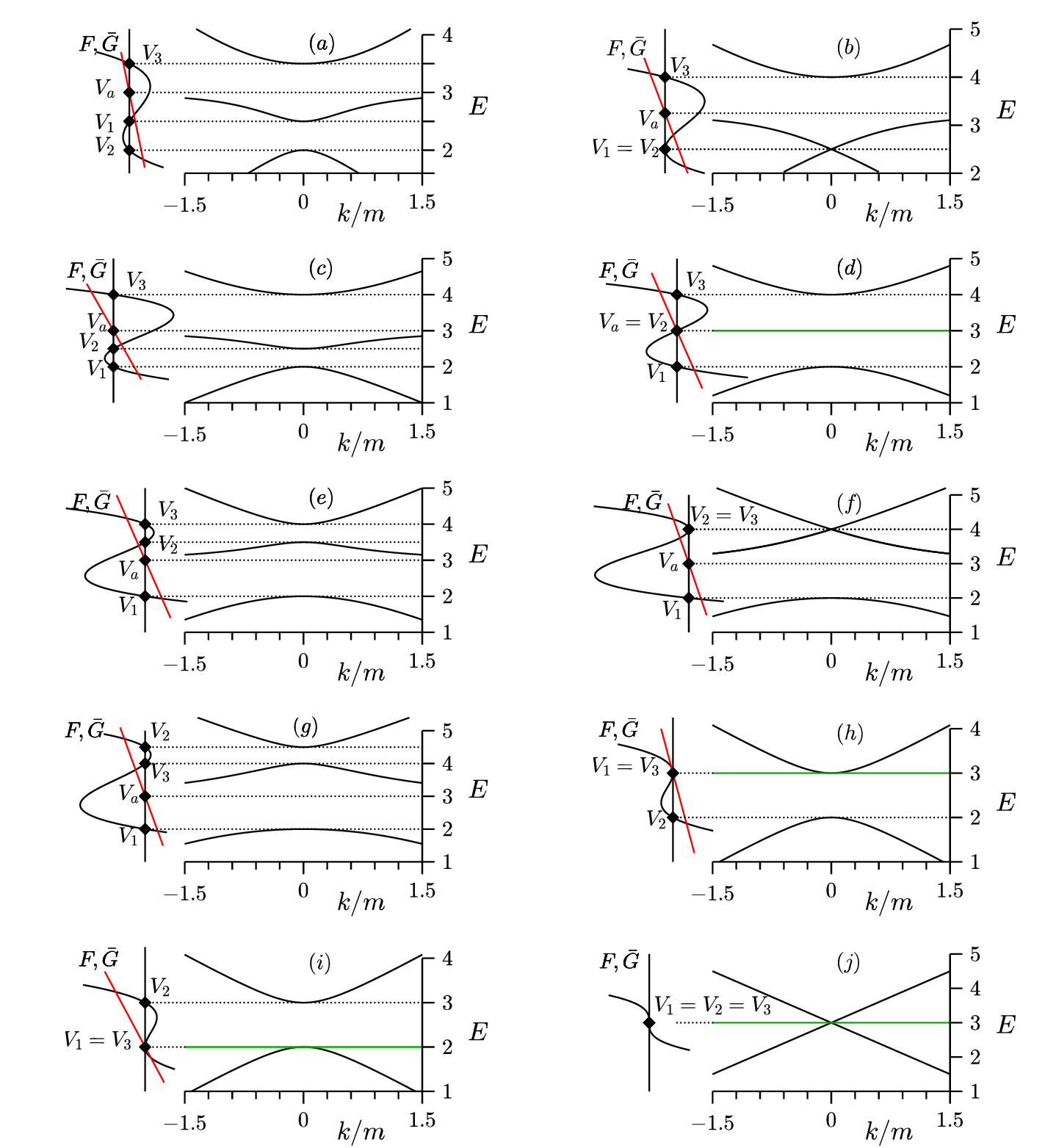}
\caption{Three (dispersive and flat) bands  for different `positions' of strength $V_2$ with respect to strengths $V_1$, $V_3$
($V_1 \le V_3$) and their middle value $V_a$\,, while successive `lifting' from bottom
to top along  $E$-axis:
 (a) $-\infty < V_2 < V_1$,  (b) $V_2 =V_1$, (c) $V_1 < V_2 < V_a$,
 (d)  $ V_2 = V_a$,  (e) $V_a < V_2 < V_3$, (f) $V_2 = V_3$, (g) $V_3 < V_2 < \infty$,
 (h) $-\infty <V_2< V_1=V_3 $, (i) $V_1=V_3 < V_2 <\infty$, (j) $V_2 = V_1=V_3$.
 Left diagrams in each panel represent both sides of
 dispersion equation (\ref{law}),  $F= F(E)$ (black curves) and
  $\bar{G} = G(E)k^2$ (red straight lines). Green horizontal lines in
  panels (d) and (h)--(j) represent flat bands.
 }
\label{fig1}
\end{figure}

For the analysis  of the three-band spectrum $E =E(k)$, where $k$ is real,
it is convenient to use  the diagrams shown in figure~\ref{fig1}.
Here, without loss of generality, it is assumed that $V_1 \le V_3$.
The solutions for the energy $E$ are indicated by the points of intersection
 of the cubic function $F(E)$ and the straight line $G(E)k^2$ that passes through
 the middle point $V_a$ located between the zeroes $E=V_1$ and $E=V_3$. The slope of
 this line is governed by $k^2 >0$ and rotating it around the `turning' point $V_a$,
 one obtains the energy bands $E(k)$. `Moving' the zero $E=V_2$ of the cubic
 function $F(E)$ along the $E$-axis,  the all possible types of the
  three-band  spectrum $E(k)$ are visually illustrated by the right diagrams
  in each panel. Each type consists of the lower and upper dispersive bands,
  whereas the middle band can be either dispersive or flat. Thus, in panels
  (a)--(c) and (e)--(g) for the case $V_1 <V_3$, we have the middle  bands of
  the dispersive  type, which are bounded. Here, in the limits
  as $V_2 \nearrow V_a$ or   $V_2 \searrow V_a$, the two-sided middle  dispersion bands shrink  to a   flat band horizontal line continuously, as  demonstrated by 
   panel (d). Finally, the case $V_1 = V_3$ is illustrated by  panels (h)--(j). 
Here,  the flat band touches the upper dispersion band if $V_2 < V_1 =V_3$,
 the lower one if $V_2 > V_1 =V_3$ and  both the lower and upper
 dispersion bands if  $V_2 =V_1 =V_3$.

 \subsection{Flat band planes}

As follows from equation (\ref{law}), the existence of flat bands is provided if
the two equalities  $F(E) =G(E)=0$ take place simultaneously. Then the dispersion law
holds true regardless of the wave number $k$. In this case, the average strength 
$V_a$ must coincide with one of the zeroes $E=V_j$\,, $j=1,\,2,\,3$, 
of cubic function (\ref{F}). Therefore, one of the three relations
\begin{equation}
  V_{1} +V_{3} =2V_j \,,~~~j=1,\,2,\,3,
 \label{flat_eqs}
\end{equation}
is the necessary and sufficient condition for the existence of flat bands.

In the case $j=2$, the corresponding relation  in (\ref{flat_eqs}) becomes
\begin{equation}
 V_{11} +V_{33}=2V_{22}\,,
 \label{rel_A}
\end{equation}
 describing a plane in the $(V_{11}\,,\, V_{22}\,,\, V_{33})$-space, which
 we call from now on the ${\cal A}$-plane. Hence, the flat band energy on this plane is
\begin{equation}
 E=V_2 =V_{22}\,.
 \label{flat_energy_A}
\end{equation}
Particularly, on the line $V_{11}= V_{22} =V_{33} \equiv V$, the flat band energy is
shifted from $E=0$ (free-particle case) to $E=V$.

In both the cases $j=1,3$, condition (\ref{flat_eqs}) reduces to one equation $V_1=V_3$\,. Consequently, this equation together with an arbitrary $V_2$, i.e.,
\begin{equation}
  V_{33}- V_{11} = 2m,~~~V_{22} \in \R,
 \label{rel_B}
\end{equation}
defines a plane in the $(V_{11}\,,\, V_{22}\,,\, V_{33})$-space,
which we call from now on the ${\cal B}$-plane. The flat band energy on this plane is
\begin{equation}
 E =V_1 =V_3= V_{11}  + m = V_{33} -m .
 \label{flat_energy_B}
\end{equation}

\begin{figure}[htb]
\includegraphics[width=0.5\textwidth,height=0.45\textwidth]{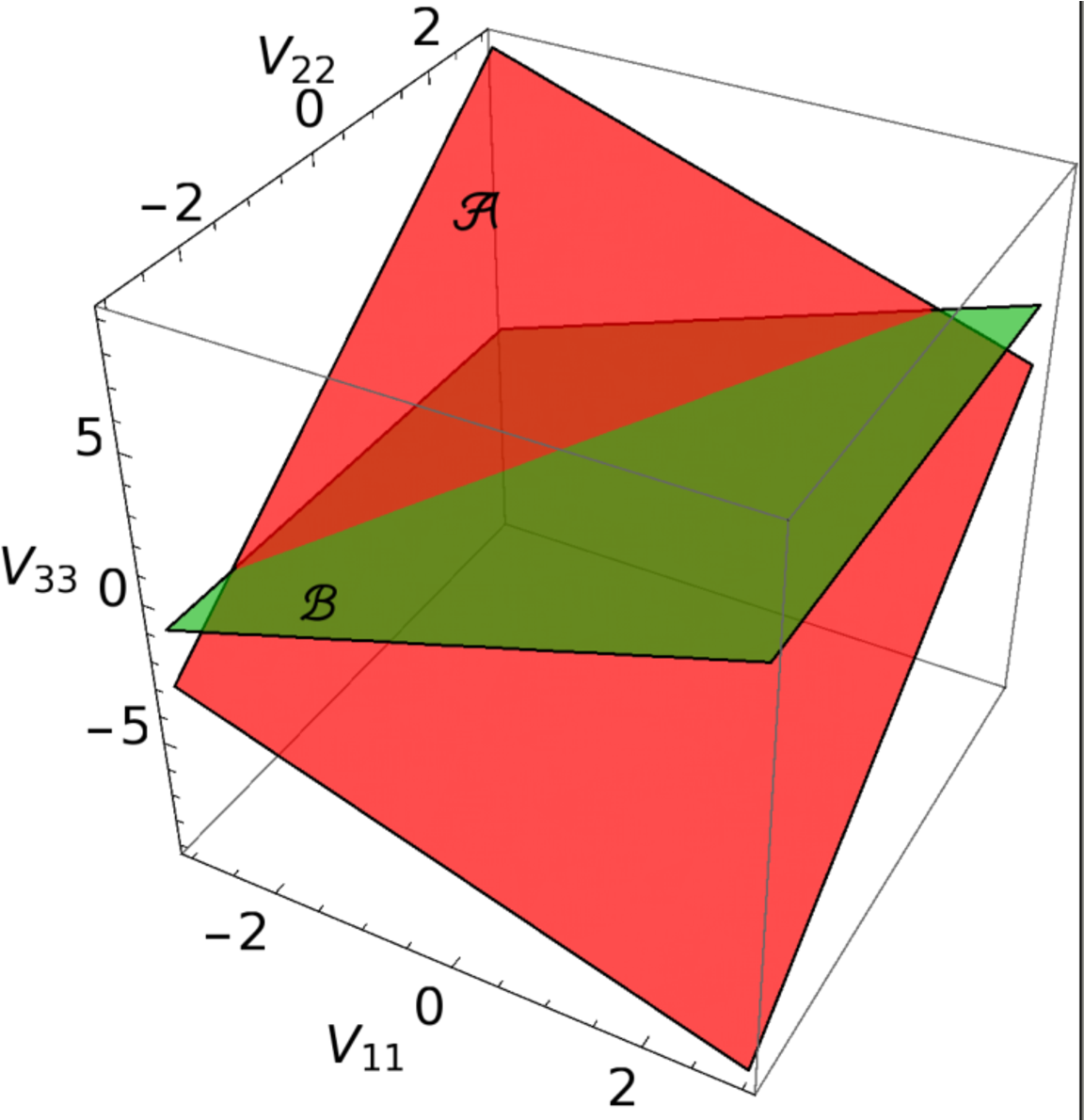}
\caption{Intersecting flat band planes ${\cal A}$ (red plane) and ${\cal B}$
(green plane).
 }
\label{fig2}
\end{figure}

In the particular case $V_1 =V_2 =V_3$\,, both equations (\ref{rel_A}) and (\ref{rel_B})
 are satisfied. Therefore, there exists an  intersection of the planes ${\cal A}$
 and ${\cal B}$ as shown in figure~\ref{fig2}.
 Consequently, on the line ${\cal A} \cap {\cal B}$, we have  the flat band energy
 \begin{equation}
 E = V_1 =V_2 =V_3=V_{11}  + m =V_{22}= V_{33} -m .
 \label{flat_energy_AB}
\end{equation}

 Thus, as illustrated by the diagrams in panels (d) and (h)--(j) of figure~\ref{fig1},
 the existence of flat bands is possible if and only if the line $G(E)k^2$ passes through
any of zeroes $E= V_j$'s, $j=1,\,2,\,3$, of the cubic function $F(E)$.
In the $(V_{11}\,,\, V_{22}\,,\, V_{33})$-space, the flat bands are found  only
on the ${\cal A}$- and ${\cal B}$-sets, including the line of their intersection
${\cal A} \cap {\cal B}$. Therefore, these sets may be called from now on as
the flat band planes. One can conclude that  panels (d) and (h)--(j) in
  figure~\ref{fig1}  illustrate all the possible types of flat bands. In other words,
  any rotation  of the straight (red) line
 around the point $E=V_a$ keeps the solution of equation (\ref{law})
 regardless of the slope determined by
 $k^2$, visualizing the existence of flat bands. Panel (d) in figure~\ref{fig1} describes
the situation as the middle dispersion curves in panels (c) and (e)  are squeezed to
a (horizontal) flat line.  The potential strengths in this case
are found on the ${\cal A}$-plane defined by equation (\ref{rel_A}) where $V_1 \neq V_3$\,.
The case $V_1 =V_3$ but $V_2 \neq V_1=V_3$ is presented by panels (h) and (i).
As illustrated by these panels,
one of the lower or upper gaps disappears in the spectrum. The potential strengths
in this case belong to the ${\cal B}$-plane defined by equation (\ref{rel_B}) where
$V_2 \neq V_1=V_3$\,. Finally, if $V_1 =V_2 =V_3$\,, the strengths are found on the
intersection of the ${\cal A}$- and ${\cal B}$-planes and, as demonstrated by
panel (j), both the gaps in the spectrum disappear.

 \subsection{Eigenenergies and eigenfunctions of dispersion bands }

 In general, if $V_1 \neq V_3$ and $V_a \neq V_j$\,, $j=1,\,2,\,3$, the situation is
 illustrated by panels (a)--(c) and (e)--(g) in figure~\ref{fig1},
 where all the three [lower $E_-(k)$, middle $E_0(k)$
 and upper $E_+(k)$]   bands  are dispersive. In this case,
 the eigenenergies  $E_\pm(k)$ and $E_0(k)$  are three roots of cubic equation (\ref{law}).
 In other cases, when the average strength $V_a$ coincides with one
 the strengths $V_j$'s,   cubic equation (\ref{law})
 reduces to a quadratic form and the triad $(V_{11},\,V_{22},\,V_{33}) \in \R^3$
 falls into one of  the ${\cal A}$- and ${\cal B}$-planes or their intersection.

 {\it Plane ${\cal A}$}: On the ${\cal A}$-plane,
owing  to relation (\ref{rel_A}),  the energy of the upper and
lower dispersion bands is a solution of the equation $(E- V_1)(E-V_3) =k^2$.
Explicitly, this solution reads
\begin{equation}
 E= E^A_\pm(k) = E^A_0  \pm \sqrt{k^2 + \left({ V_{1} -V_{3}\over 2}
 \right)^{\!\!2} } \,, \quad E^A_0 =V_2\,,
 \label{E(k)_A}
 \end{equation}
illustrated by  panel (d) in figure~\ref{fig1}, where the energies of the
lower and upper dispersion bands
 $E^A_\pm(k)$ are depicted by the black curves and the energy of the  flat band $E^A_0$
 by the blue horizontal line. Both the lower ($V_1 < E < E_0^A$) and upper
 ($E_0^A < E< V_3$) gaps are non-empty.
 The corresponding three eigenfunctions $\psi_\pm^A(k;x)$ and
$\psi_0^A(k;x)$ are described by general solution  (\ref{gen_sol}), where
$\sigma_1$ and $\sigma_3$ are substituted by
\begin{eqnarray}
 &&\!\!\!\!\!\!\!\!\!\!\!\!\!\!\!\!\!\!\!\!\!\!\!\!\!\!
 \sigma^A_{1,\pm}(k)={{\rm i} \over \sqrt{2} \, k}\left[{ V_{1} -V_3 \over 2}
 \pm \sqrt{k^2 + \left( { V_1 -V_3 \over 2} \right)^{\!\!2} }\,\right],~~
 \sigma_{1,0}^A(k) ={{\rm i} \over \sqrt{2} \, k}(V_2 -V_3) ,
 \nonumber \\ \label{sigma_1_3_A} \\
&&\!\!\!\!\!\!\!\!\!\!\!\!\!\!\!\!\!\!\!\!\!\!\!\!\!\!
\sigma^A_{3,\pm}(k)={{\rm i} \over \sqrt{2} \, k}\left[{ V_{3} -V_1 \over 2}
 \pm \sqrt{k^2 + \left( { V_1 -V_3 \over 2}
 \right)^{\!\!2} }\,\right],~~\sigma_{3,0}^A(k) ={{\rm i}
 \over \sqrt{2} \, k}(V_2 -V_1) ,
 \nonumber
\end{eqnarray}
respectively.

{\it Plane ${\cal B}$}:
Similarly, on the ${\cal B}$-plane, we arrive at
the equation $(E-V_1)(E-V_{2}) =k^2$, having the solution
\begin{equation}
E= E^B_\pm(k) ={ V_1 +V_2 \over 2}  \pm \sqrt{k^2 + \left( {V_{1} -V_{2} \over 2}
 \right)^{\!\!2} } \, ,\quad E_0^B = V_1 =V_3\,.
 \label{E(k)_B}
\end{equation}
 This solution is depicted  in figure~\ref{fig1}  for two configurations, where
 the strength $V_2$ does not coincide with the flat band energy $E_0^B$:
 in panel (h) $V_2 < E_0^B$ and in panel (i) $V_2> E_0^B$. Correspondingly,
 only the lower gap $V_2 < E< E_0^B$ and the upper gap $E_0^B < E< V_2$  are non-empty,
 while respectively the upper and lower gaps disappear. Similarly, the eigenfunctions
 $\psi_\pm^B(k;x)$ and $\psi_0^B(k;x)$ are given by wave function (\ref{gen_sol}),
 where $\sigma_1$ and $\sigma_3$ are replaced by
\begin{eqnarray}
&&\sigma^B_{1,\pm}(k)= \sigma^B_{3,\pm}(k)={{\rm i} \over \sqrt{2} \, k}
\left[{ V_{1} -V_2 \over 2}  \pm \sqrt{k^2 + \left( { V_1 -V_2 \over 2}
 \right)^{\!\!2} }\,\right] , \nonumber \\ \label{sigma_1_3_B} \\
 && \sigma^B_{1,0}(k) =\sigma^B_{3,0}(k) ={{\rm i} \over \sqrt{2} \, k}(V_1-V_2)=
 {{\rm i} \over \sqrt{2} \, k}(V_3-V_2).
\nonumber
\end{eqnarray}

{\it Line ${\cal A}\cap {\cal B}$}: On the ${\cal A}\cap{\cal B}$-line,
energies (\ref{E(k)_A}) and (\ref{E(k)_B}) reduce to
\begin{equation}
E= E_\pm^{A\cap B}(k) =E_0^{A\cap B} \pm |k|, \quad E_0^{A\cap B} = V_1 =V_2=V_3\,.
 \label{E(k)_A_B}
\end{equation}
As illustrated by panel (j) in figure~\ref{fig1}, both the gaps for this configuration disappear.
Setting in (\ref{sigma_1_3_A}) and (\ref{sigma_1_3_B}) $V_1=V_2 =V_3$, we obtain
\begin{equation}
\!\!\!\!\!\!\!
 \sigma^{A \cap B}_{1,\pm}(k) = \sigma^{A \cap B}_{3,\pm}(k) =
 \pm\, {\rm sgn}(k){ {\rm i} \over \sqrt{2}}\, , \quad
 \sigma^{A \cap B}_{1,0}(k) = \sigma^{A \cap B}_{3,0}(k) =0.~~
\label{sigma_1_3_A_B}
\end{equation}

Finally, notice that, as follows from the general formula for energy (\ref{E(k)_A}), in the particular case  $V_{11} = V_{22} =V_{33} \equiv V$, we have the energy shift
$E =V \pm \sqrt{k^2 +m^2} $ from the free-particle spectrum, similarly to
the one-dimensional non-relativistic case. The components of the wave function,
in this case, are as follows
\begin{eqnarray}
\!\!\!\!\!\!\!\!\!\!\!\!
\hspace{-5pt} \psi_+(k;x)&=&B_1 \left( \begin{array}{ccc} m+\sqrt{k^2 +m^2}\\
{\rm i}\sqrt{2}\,k\\
  m-\sqrt{k^2 +m^2} \end{array} \right){\rm e}^{{\rm i}kx} +
B_2 \left( \begin{array}{ccc} m+\sqrt{k^2 +m^2}\\ - {\rm i}\sqrt{2}\,k\\
  m-\sqrt{k^2 +m^2} \end{array} \right){\rm e}^{-{\rm i}kx}, \nonumber \\
  \!\!\!\!\!\!\!\!\!\!\!
\hspace{-5pt}   \psi_0(k;x)&=&B_1 \left( \begin{array}{ccc} m \\ {\rm i}\sqrt{2}\,k\\
  m \end{array} \right){\rm e}^{{\rm i}kx} +
B_2 \left( \begin{array}{ccc} m\\ - {\rm i}\sqrt{2}\,k\\
  m \end{array} \right){\rm e}^{-{\rm i}kx}, \label{psi_pm_0}\\
  \!\!\!\!\!\!\!\!\!\!\!
\hspace{-5pt}   \psi_-(k;x)&=&B_1 \left( \begin{array}{ccc} m-\sqrt{k^2 +m^2}\\ {\rm i}\sqrt{2}\,k\\
  m+\sqrt{k^2 +m^2} \end{array} \right){\rm e}^{{\rm i}kx} +
B_2 \left( \begin{array}{ccc} m-\sqrt{k^2 +m^2}\\ - {\rm i}\sqrt{2}\,k\\
  m+\sqrt{k^2 +m^2} \end{array} \right){\rm e}^{-{\rm i}kx} , \nonumber
\end{eqnarray}
where $B_1$ and $B_2$ are arbitrary constants. Particularly, on the line
${\cal A}\cap {\cal B}$, where the gapless spectrum occurs,  wave function components
(\ref{psi_pm_0}) are simplified to  the form
\begin{eqnarray}
\hspace{-10pt}  \psi_+(k;x)\!\!&=&\!\! B_1 \!\left( \begin{array}{ccc} 1 \\
{\rm sgn}(k)\,  {\rm i}\sqrt{2}\\
  -1 \end{array} \right){\rm e}^{{\rm i}kx} +
B_2 \! \left( \begin{array}{ccc} 1 \\ -{\rm sgn}(k)\, {\rm i}\sqrt{2}\\
  -1 \end{array} \right){\rm e}^{-{\rm i}kx} , \nonumber \\
\hspace{-10pt}   \psi_0(k;x)\!\!&=&\!\! B_1 \! \left( \begin{array}{ccc} 0 \\
  {\rm sgn}(k)\,{\rm i}\sqrt{2}\\
  0 \end{array} \right){\rm e}^{{\rm i}kx} +
B_2\! \left( \begin{array}{ccc} 0\\ -{\rm sgn}(k) \,{\rm i}\sqrt{2}\\
  0 \end{array} \right){\rm e}^{-{\rm i}kx},  \\
\hspace{-10pt}   \psi_-(k;x)\!\!&=&\!\!B_1\! \left( \begin{array}{ccc} -1
  \\ {\rm sgn}(k)\,{\rm i}\sqrt{2}\\
  1 \end{array} \right){\rm e}^{{\rm i}kx} +
B_2\! \left( \begin{array}{ccc} -1\\ -{\rm sgn}(k) \,{\rm i}\sqrt{2}\\
  1 \end{array} \right){\rm e}^{-{\rm i}kx}, \nonumber
\end{eqnarray}
with arbitrary constants $B_1$ and $B_2$\,.

\section{Bound states of the Hamiltonian with rectangular potentials}

In the previous section we have examined the solutions of system (\ref{basic_eqs})
with a  three-component potential $V(x) =(V_{11},\, V_{22}, \,V_{33})$, which is  constant on the whole $x$-axis. Bound states can be materialized if the  potential
$V(x)$ is compactly supported on some finite interval.
In this regard, we focus here on the potential components, each being of 
a rectangular form. More precisely, we assume
\begin{equation}
 V(x)= \left\{ \begin{array}{ll} \smallskip
 {\rm col}(V_{11},\, V_{22}, \,V_{33}) & ~\mbox{for}~
                x_1 \le x \le x_2\,, \\
                {\rm col}(0,\,0,\,0) & ~\mbox{for}~
                (-\infty,\,x_1)\cup (x_2,\,\infty),
               \end{array} \right.
\label{V(x)}
\end{equation}
where the points $x_1$ and $x_2$ are arbitrary.

Within the interval $x_1 < x < x_2$\,, the representation of general solution
(\ref{gen_sol}) can also be rewritten in the terms of trigonometric functions as follows
\begin{equation}
\!\!\!\!\!\!\!\!\!\!\!\!\!\!\!\!\!\!\!\!\!\!\!\!\!\!
  \psi(x) =C_1 \left(
  \begin{array}{ccc} \smallskip ( E -V_1)^{-1}k  \sin(kx)  \\ \smallskip
  \sqrt{2}\, \cos(kx) \\ \smallskip
 (V_3 -E)^{-1}k \sin(kx)  \end{array}  \right) +
 C_2 \left(   \begin{array}{ccc} \smallskip ( E -V_1)^{-1}k  \cos(kx)  \\ \smallskip
 - \sqrt{2}\, \sin(kx) \\ \smallskip
 (V_3 -E)^{-1}k \cos(kx)  \end{array}  \right) , ~~
 \label{sol_in}
\end{equation}
where $C_1$ and $C_2$ are arbitrary constants. For realizing bound states,
beyond the interval $x_1 < x < x_2$, wave function (\ref{gen_sol}) must decrease
to zero at infinity.  Setting  in (\ref{gen_sol})--(\ref{k}),
 $k= {\rm i}\kappa$, $\kappa >0$, $V_1=m$, $V_2=0$ and $V_3 =-m$,
  we arrive at the following finite representation
of general solution (\ref{gen_sol}) outside the interval $x_1 < x < x_2$:
\begin{equation}
 \psi(x) = \left\{ \begin{array}{ll} \smallskip
 D_1\, {\rm col}(\rho^{-1},\, \sqrt{2}, \,\,\rho)\,
 {\rm e}^{\kappa (x-x_1)} & \mbox{for}~- \infty <x \le x_1\,, \\
  D_2\, {\rm col}( \rho^{-1}, \, - \sqrt{2},\, \, \rho)\,
 {\rm e}^{- \kappa (x-x_2)} & \mbox{for}~ ~x_2 \le x < \infty ,
\end{array} \right.
\label{sol_out}
\end{equation}
where $D_1$ and $D_2$ are arbitrary constants,
\begin{equation}
 \kappa : = \sqrt{m^2-E^2} \quad \mbox{and} \quad
 \rho := \sqrt{m-E \over  m+E}\,.
 \label{kappa}
\end{equation}
The four constants $C_1$\,, $C_2$ and $D_1$\,, $D_2$ in expressions (\ref{sol_in})
and (\ref{sol_out}) can be determined by using matching conditions imposed
on the boundaries $x=x_1$ and $x=x_2$\,. The requirement for continuity of all the
three components of the wave function $\psi(x)$
at $x=x_1$ and $x=x_2$ leads to  six equations that involve only the four constants.
Therefore such matching conditions are not appropriate. However, as can be seen from
the structure of system  (\ref{basic_eqs}), it is not necessary to require the continuity
of the components $\psi_1(x)$ and $\psi_3(x)$. Instead, it is sufficient to impose
the continuity of  $\psi_1(x) -\psi_3(x)$ and $\psi_2(x)$, so that the components
$\psi_1(x)$ and $\psi_3(x)$, each alone, may in general be discontinuous at $x_1$
and $x_2$.
Thus, from (\ref{sol_in}) and (\ref{sol_out}), we  obtain the following four equations:
\begin{equation}
 \begin{array}{rrrr} \smallskip  C_1 \sin(kx_1) +C_2 \cos(kx_1)
 & =& D_1/\gamma , \\ \smallskip
   C_1 \cos(kx_1) - C_2 \sin(kx_1)   &=& D_1\,, \\ \smallskip
   C_1 \sin(kx_2) +C_2 \cos(kx_2)   &=& D_2/\gamma , \\ \smallskip
    C_1 \cos(kx_2) -C_2 \sin(kx_2)   &=& - D_2\,,
   \end{array}
 \label{matching}
\end{equation}
where $k$ is given by (\ref{k}) and
\begin{equation}
  \gamma :=  {\kappa \over k}\left(1 -{V_2 \over  E}\right).
 \label{gamma}
\end{equation}
Note that the boundary conditions, imposed above on the components
$\psi_1(x) - \psi_3(x)$ and $\psi_2(x)$ at $x=x_1$ and $x=x_2$,
provide the continuity of the net current
\begin{equation}
j(x)=\psi^\dag S_y\psi ={{\rm i} \over \sqrt{2}}
[\psi_2^{*}(\psi_1-\psi_3)-(\psi_1^{*}-\psi_3^{*})\psi_2].
\label{current}
\end{equation} 

Equating the determinant of the system of equations (\ref{matching})
to zero, one can derive a necessary condition for the existence of bound states.
In general, the solution inside the interval $x_1 \le x \le x_2$
 can be given through  a  matrix  connecting the values of the functions
 $\psi_1(x) -\psi_3(x)$ and $\psi_2(x)$ at  the boundary  points $x=x_1$ and $x=x_2$.
 We define this connection matrix as follows
 \begin{equation}
\!\!\!\!\!\!\!\!\!\!
 \left( \begin{array}{cc} (\psi_1-\psi_3)(x_2) \\ \psi_2(x_2) \end{array} \right)=
 \Lambda \left( \begin{array}{cc} (\psi_1-\psi_3)(x_1) \\ \psi_2(x_1) \end{array} \right),
 ~~\Lambda := \left( \begin{array}{ll} \lambda_{11}~~\lambda_{12} \\
        \lambda_{21}~~\lambda_{22}   \end{array} \right).~~~
        \label{L}
\end{equation}
Using then the boundary values of the components  $\psi_1(x) -\psi_3(x)$ and $\psi_2(x)$ obtained from wave function (\ref{sol_out})  and excluding the constants
$D_1$ and $D_2$, we get the equation for the bound state energy $E=E_b$ given
in terms of
the connection matrix $\Lambda$ that describes any potential profile inside the interval $x_1 \le x \le x_2$:
\begin{equation}
 \lambda_{11}+ \lambda_{22} + {\kappa \over \sqrt{2}\,E}\lambda_{12}
 + {\sqrt{2}\,E \over \kappa }\lambda_{21} =0,
 \label{gen_bs_eq}
\end{equation}
where $\kappa$ is defined in (\ref{kappa}). In one dimension, similar equations have been established  in \cite{ZZ2014,ZZ2021} for the non-relativistic Schr\"{o}dinger equation and in  \cite{Gusynin2022} for the Dirac equation.

\subsection{Explicit formula for the connection matrix $\Lambda$}

In the particular case of solution (\ref{gen_sol}), the connection matrix $\Lambda$
 can be calculated explicitly.  Indeed, using this solution on the interval
 $x_1 \le x \le x_2$, we write
\begin{equation}
 \begin{array}{ll} \smallskip
  (\psi_1 -\psi_3)(x) &= {\rm i}\eta \left( -B_1{\rm e}^{{\rm i}kx} +
  B_2{\rm e}^{-{\rm i}kx} \right), \\
 ~~~~~~~\psi_2(x) &= B_1{\rm e}^{{\rm i}kx} +   B_2{\rm e}^{-{\rm i}kx},
 \end{array}
 \label{system_in}
\end{equation}
with
\begin{equation}
 \eta := -{\rm i}(\sigma_1 +\sigma_3)= {\sqrt{2} \over k}(E-V_2) ,
 \label{eta}
\end{equation}
where $k$ is given by (\ref{k}).
Fixing equations (\ref{system_in}) at $x =x_1$, we find from these equations the
constants $B_1$ and $B_2$ and then substitute these values again into
equations (\ref{system_in}), but now fixed at $x=x_2$. As a result, we get the
$\Lambda$-matrix in the form
\begin{equation}
 \Lambda  =\left( \begin{array}{cc}         ~~~ \cos(kl)~~~~~~~\eta \sin(kl) \\
        -\eta^{-1}\sin(kl)~~~\cos(kl)   \end{array} \right), \quad l := x_2 -x_1\,,
        \label{L_matrix}
\end{equation}
where  $k$ and $\eta$ are given by formulas (\ref{k}) and (\ref{eta}), respectively.

\subsection{Basic equations for bound state energies}

Inserting the elements of $\Lambda$-matrix (\ref{L_matrix}) into general equation
(\ref{gen_bs_eq}), we obtain the equation for the energy of bound states $E=E_b$
in the form
\begin{equation}
 2 + \left( \gamma - {1 \over \gamma}  \right) \tan(kl)=0,
 \label{bs_eq}
\end{equation}
where the functions $k=k(E)$ and $\gamma(E)$ are defined by formulas (\ref{k}) and
(\ref{gamma}), respectively.

Equation (\ref{bs_eq}) splits into two  simple equations with respect to the
unknowns, which we denote from now on as $E =E^+$ and $E= E^-$. As a result,
these equations read
\begin{equation}
 \gamma= \left\{ \smallskip \begin{array}{ll}
 -\cot(kl/ 2)      & \mbox{for}~~ E =E^+ \,, \\ ~~ \tan(kl/ 2) &
 \mbox{for}~~ E =E^- \,.  \end{array} \right.
 \label{bs_eqs}
\end{equation}
The solutions to these equations, where $k(E)$ and $\gamma(E)$ are given
by (\ref{k}) and (\ref{gamma}),  describe  bound state energies $E= E^\pm_b$,
the total number of which at a given three-component strength
 $(V_{11},\,V_{22},\,V_{33}) \in \R^3$ may be finite or even infinite.
Each of these energies must belong to the gap $(-m,m)$.
The existence of the solutions  $E^\pm_b \in (-m, m)$  follows from argument
that each of equations (\ref{bs_eqs}) can be represented in the form 
$\kappa /E = f(E)$,
where the function $f(E)$ varies on the interval $-m < E < m$  slowly than 
$\kappa/E$.

\subsection{Bound state eigenfunctions}

From matching conditions (\ref{matching}), one can write the relations between
the constants $C_1$ and $C_2$ as follows
\begin{eqnarray}
 C_1\left[\cos(kx_1)-\gamma \sin(kx_1)\right] &= & C_2\left[ \sin(kx_1)+\gamma \cos(kx_1)\right],
 \nonumber \\ C_1\left[ \cos(kx_2)+\gamma \sin(kx_2)\right] &=&
 C_2\left[ \sin(kx_2)- \gamma \cos(kx_2)\right],  \nonumber
\end{eqnarray}
which are equivalent because of equation (\ref{bs_eq}).
Inserting here $\gamma$ from equations (\ref{bs_eqs}), we find
\begin{equation}
\begin{array}{ll} \smallskip  C_1 \sin(ka) = -C_2  \cos(ka) & \mbox{for}~E^+ ,\\
  C_1 \cos(ka) =~~C_2 \sin(ka) & \mbox{for}~E^-,
         \end{array} \quad a := \frac{1}{2}(x_1 +x_2).
         \label{C_1_C_2}
\end{equation}
Using next equations (\ref{bs_eqs}) and relations  (\ref{C_1_C_2}) in general 
solution (\ref{gen_sol}),
 we obtain on the interval $x_1 < x < x_2$  the following two (even and odd parity)
 forms for the wave function:
\begin{eqnarray}
\!\!\!\!\!\!\!\!\!\!\!\!\!\!\!
\psi^+(x) &= &{C_1 \over \cos(ka)} \left(
  \begin{array}{lll} \smallskip  (E -V_1)^{-1}k \sin[k(x-a)]  \\ \smallskip
  \sqrt{2}\, \cos[k(x-a)] \\ \smallskip
 (V_3 -E)^{-1} k \sin[k(x-a)]  \end{array}  \right) ~~\mbox{for}~~E=E^+, ~~
 \label{psi+_in} \\
 \!\!\!\!\!\!\!\!\!\!\!\!\!\!\!
 \psi^-(x) &= & {C_2 \over \cos(ka)}\left(
  \begin{array}{lll} \smallskip ( E -V_1)^{-1}k \cos[k(x-a)]  \\ \smallskip
 - \sqrt{2}\, \sin[k(x-a)] \\ \smallskip
 ( V_3 -E)^{-1} k \cos[k(x-a)]  \end{array}  \right)~~ \mbox{for}~~E=E^-.~~
          \label{psi-_in}
\end{eqnarray}
The parity transformation of a three-component fermion $\psi(x)$ is defined as
the reflection with respect to a point $x=a$: $\psi(x+a)
\rightarrow \psi^P(x)=P \psi(-x+a)$, where the matrix $P={\rm diag}(-1,1,-1)$
anti-commutes with $S_y$ and commutes with $S_z$.

\begin{figure}[htb]
\begin{centering}
\includegraphics[width=0.85\textwidth]{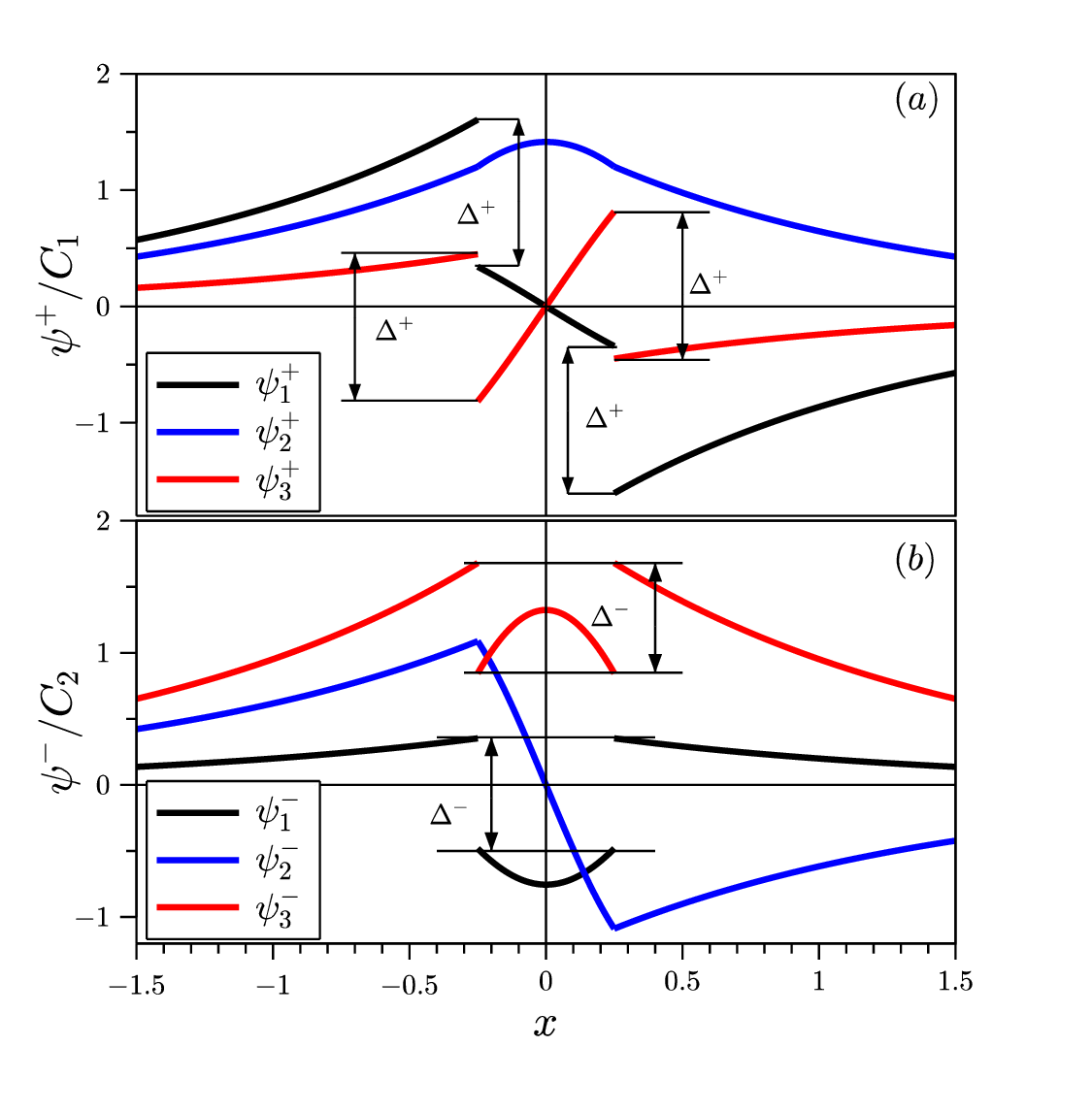}
\caption{Wave functions $\psi^\pm(x)$ for the potential
 with strengths $V_{11}= V_{22}= V_{33} \equiv V= 3\,m$ and $l=0.5\,m^{-1}$
 plotted according to formulas (\ref{psi+_in})--(\ref{psi-_out}).
For this potential, the solutions to equations (\ref{bs_eqs}) are
$E^+=  0.56\,m$ and $E^- = - 0.65\,m$.  Space $x$ is measured in units of $m^{-1}$.
 }
\label{fig3}
\end{centering}
\end{figure}

Beyond the interval $x_1 \le x \le x_2$\,,
 from matching conditions (\ref{matching}), one can find the constants
$D_1$ and $D_2$.  Using then relations (\ref{C_1_C_2}), we  get the wave functions
$\psi^\pm(x)$ in the form
\begin{equation}
\!\!\!\!\!\!\!\!\!\!\!\!\!\!\!\!\!\!\!\!\!\!\!\!\!\!
 \psi^{+}(x) =C_1 {\cos(kl/2) \over \cos(ka)}
 \left\{ \begin{array}{ll} \smallskip
  {\rm col}\!\left(\rho^{-1}, \, \sqrt{2}\,, \, \rho
 \right)\!{\rm e}^{\kappa (x-x_1)}, & -\infty < x < x_1\,, \\
  {\rm col}\!\left(-\rho^{-1}, \,  \sqrt{2}\,, \,- \rho
 \right)\!{\rm e}^{-\kappa (x-x_2)}, & ~x_2 < x < \infty \,, \end{array} \right.
 \label{psi+_out}
\end{equation}
for $E=E^+$ and
\begin{equation}
\!\!\!\!\!\!\!\!\!\!\!\!\!\!\!\!\!\!\!\!\!\!\!\!\!\!
 \psi^{-}(x) = C_2 {\sin(kl/2) \over \cos(ka)}
 \left\{ \begin{array}{ll} \smallskip
  {\rm col}\!\left(\rho^{-1}, \, \sqrt{2}\,, \, \rho
 \right)\!{\rm e}^{\kappa (x-x_1)}, & -\infty < x < x_1\,, \\
  {\rm col}\!\left(\rho^{-1}, \, - \sqrt{2}\,, \, \rho
 \right)\!{\rm e}^{-\kappa (x-x_2)}, & ~x_2 < x < \infty \,, \end{array} \right.
 \label{psi-_out}
\end{equation}
for $E=E^-$. Note that representation (\ref{psi+_in})--(\ref{psi-_out}) allows us
to set here $a=0$. Particularly, $a=0$ if $x_1 =-l/2$ and $x_2 =l/2$.
The shape of the eigenfunctions $\psi^\pm(x)$
 given by formulas (\ref{psi+_in})--(\ref{psi-_out})  is illustrated by figure~\ref{fig3}.
Notice that  the discontinuity of the components $\psi_1(x)$ and $\psi_3(x)$
at the boundaries $x=x_1$ and $x=x_2$ is calculated as follows
\begin{equation}
\!\!\!\!\!\!\!\!\!\!\!\!\!
 \begin{array}{ll} \smallskip
 \psi^+_j(x_1-0)-\psi^+_j(x_1+0)=\psi^+_j(x_2-0)-\psi^+_j(x_2+0)=C_1\Delta^+,  \\
 \psi^-_j(x_1-0)-\psi^-_j(x_1+0)=\psi^-_j(x_2+0)-\psi^-_j(x_2-0)=C_2\Delta^- ,
 \end{array}
\end{equation}
where $j=1,\,3$ and
\begin{equation}
\!\!\!\!\!\!\!\!\!\!\!\!\!\!\!\!\!\!\!\!\!\!\!\!\!\!\!\!\!
 \Delta^\pm = {\mu \over \kappa \cos(ka)} \left\{ \begin{array}{ll} \smallskip
 \cos(kl/2) & \mbox{for}~E= E^+,  \\
 \sin(kl/2) & \mbox{for}~E= E^-, \end{array} \right. \;\;
 \mu : = m - { E\,(V_1 - V_3) \over 2E- V_1 -V_3 }\, .
\end{equation}
In the particular case with $V_{11} =V_{33} \equiv 0$
and an arbitrary $V_{22}$, we have $\mu =0$
for any energy $E$, so that $\psi_1(x)$ and $\psi_3(x)$ are continuous at
$x_1$ and $x_2$ in this particular case.

\section{Characteristic spectra of bound states}

Based on equations  (\ref{bs_eqs}), where $k$ and $\gamma$ are given by
expressions (\ref{k}) and (\ref{gamma}),
a whole variety of bound states can be materialized,
regarding various values  of the rectangular potential strengths
in formula (\ref{k}), where $k$ may be either real or imaginary. In particular, on
the  flat bands  defined by equations (\ref{rel_A}) and (\ref{rel_B}),
 expression (\ref{k}) is simplified, reducing to the equations
\begin{equation}
 k= \left\{ \begin{array}{lll} \smallskip
 \sqrt{(E-V_1)(E- V_3)} & ~~(V_{2}=V_a) & \mbox{for}~{\cal A},\\
   \smallskip   \sqrt{(E-V_1)(E- V_{2})} & ~~(V_{1} = V_3) & \mbox{for}~{\cal B}, \\
 |E-V_{2}| & ~~(V_{1}=V_{2}=V_{3})  & \mbox{for}~{\cal A}\cap{\cal B}.
 \end{array} \right.
 \label{k_A_B}
\end{equation}
However, the bound states can also exist if the strengths are found beyond the flat band planes ${\cal A}$ and ${\cal B}$. For the following analysis of the 
existence of bound
states, we restrict ourselves to the investigation on two pencils of straight lines 
in  the $(V_{11},\,V_{22},\, V_{33} )$-space using a single strength parameter $V$.
 In general, a pencil of lines is defined as the set of lines passing through
 a common point (vertex) in the space. We have chosen two such points in
 the $(V_{11},\,V_{22},\, V_{33} )$-space:  $(0,\,0\,0)$ and $(-m,\,0,\,m)$.
More precisely, we consider the following two pencil representations:
\begin{equation}
V_{11}= \alpha_1V,~~V_{22}= \alpha_2V, ~~V_{33}= \alpha_3V ,
\label{rep1}
\end{equation}
with the vertex at the origin $(0,\,0,\,0)$ and
\begin{equation}
V_{1}= \alpha_1V ,~~~V_{2}= \alpha_2V, ~~~V_{3}= \alpha_3V ,
\label{rep2}
\end{equation}
with the vertex at the shifted point $(-m,\,0,\,m)$.
Here, $\alpha_j \in \R$, $j=1,\,2,\,3$, with certain  constraints 
to be imposed below
in each particular  case. In the following, we refer to these representations as to
the pencils $P_1$ and $P_2$, respectively.
Correspondingly, wave number (\ref{k}) takes the following forms:
\begin{equation}
\!\!\!\!\!\!\!\!\!\!\!\!\!\!\!\!\!\!\!\!\!\!\!\!\!
k = \sqrt{2(E-\alpha_1V-m)(E-\alpha_2V)(E-\alpha_3V +m) \over 2E -
(\alpha_1 + \alpha_3)V} \quad \mbox{for pencil}~P_1
\label{kP1}
\end{equation}
and
\begin{equation}
k = \sqrt{2(E-\alpha_1V)(E-\alpha_2V)(E-\alpha_3V ) \over 2E -
(\alpha_1 + \alpha_3)V} \quad \mbox{for pencil}~P_2\,.
\label{kP2}
\end{equation}

Some of the lines from the pencils $P_1$ and $P_2$ fall into the flat band planes
${\cal A}$ and
${\cal B}$. For example, the line with $\alpha_1 =\alpha_2= \alpha_3 $ from
the pencil $P_1$
 corresponds to the  potential referred  in \cite{Zhang2022JPB} as the potential of
 type I and the corresponding
line falls into the ${\cal A}$-plane. The other two particular cases of the pencil $P_1$
are $\alpha_1 =\alpha_3 =0$, $\alpha_2 \neq 0$  (type II, as referred in
\cite{Zhang2022JPB}) and
$\alpha_1 \neq 0$, $\alpha_2 =\alpha_3 =0$ (type III, as defined in \cite{Zhang2022JPB}).
Both these examples correspond to the potentials with  strengths found outside
the flat band planes. The lines with $\alpha_1 = \alpha_3 $ ($V_1 =V_3$)
from the pencil $P_2$ fall into the
${\cal B}$-plane. Finally, the particular example  $\alpha_1 =\alpha_2= \alpha_3 $
($V_1 =V_2 =V_3$) in the pencil $P_2$ corresponds to the ${\cal A}\cap {\cal B}$-line.

Solving equations (\ref{bs_eqs}) with $\gamma$ and $k$ given by (\ref{gamma}),
(\ref{kP1}) and (\ref{kP2}), for admissible fixed values of the coefficients
$\alpha_j$'s, one can investigate  a bound state energy $E=E_b$ (if any)
 as a function of the strength $V$ on the whole $V$-axis. As demonstrated below,
 different scenarios of such a behavior occur
that depend on the configuration of $\alpha_j$'s in the pencils $P_1$ and $P_2$.
Thus, the  number of bound states at a given value of $V$ may be finite or even infinite.
For some configurations of the coefficients $\alpha_j$'s, the number of bound
states may increase owing to detachments from the thresholds $E= \pm m$.
 Notice that, due to equations (\ref{sol_out}) where $\kappa >0$,
for any configuration of potential strengths, the energy $E_b$ must be
  found in the gap $(-m,m)$.

\subsection{Bound states with asymptotically periodic energy behavior }

Here we introduce the notion `asymptotic periodicity', which  means that the solutions
to equations (\ref{bs_eqs}), consisting  of repeating pieces on the $V$-axis,
in the limit as $|V| \to \infty$,  become exactly periodic.
Such a behavior can occur if $\gamma \to $ const.$\,\neq 0$ and $k \propto |V|$
for large $V$. This happens if all the coefficients $\alpha_j$'s
in both representations (\ref{rep1}) and (\ref{rep2}) are non-zero.
  Without loss of generality, one can put here $\alpha_2 =1$.

For large $V$, the asymptotic representation of equations (\ref{bs_eqs}) can be treated
as follows.  For both the pencils $P_1$ and $P_2$, according to
(\ref{gamma}), (\ref{kP1}) and (\ref{kP2}), we have $k \sim \sqrt{\beta}\,|V|$ and
$\gamma \sim - {\rm sgn}(V) \kappa/\sqrt{\beta}\,E$ where
\begin{equation}
\beta := {2\alpha_1 \alpha_3 \over \alpha_1 +\alpha_3} \in \R \quad
(\alpha_1+ \alpha_3 \neq 0),
\label{beta}
\end{equation}
so that asymptotically equations (\ref{bs_eqs}) become
\begin{equation}
 {\kappa \over \sqrt{\beta}\,E} \sim
  \left\{ \begin{array}{ll} \smallskip ~~\cot\!\left(\sqrt{\beta}\,Vl/2\right) &
  \mbox{for}~E=E^+, \\
- \tan\!\left(\sqrt{\beta}\,Vl/2\right) & \mbox{for}~E=E^-.  \end{array} \right.
\label{bs_P_eqs}
 \end{equation}
From these asymptotic relations, for $\beta >0$, we get the periodic behavior:
\begin{equation}
\!\!\!\!\!\!\!\!\!\!\!\!\!\!\!\!\!\!\!\!\!
\left(\begin{array}{ll} \smallskip  E^+_b \\ E^-_b \end{array} \right)
\simeq {\rm sgn}\!\left(\!\tan{\sqrt{\beta}\,Vl \over 2}\right)m
\left( \begin{array}{ll}
\smallskip ~~\left[ 1 + \beta \cot^2(\sqrt{\beta}\,Vl/2)\right]^{-1/2} \\
- \left[ 1+ \beta\tan^2(\sqrt{\beta}\,Vl/2)\right]^{-1/2}  \end{array} \right)
\label{E_P}
\end{equation}
that  confirms the asymptotic periodicity of the bound state energies $E^\pm_b$ for
both the  pencils $P_1$ and $P_2$. In the particular case $\beta=1$, solutions
(\ref{E_P}) are simplified reducing to the form
\begin{equation}
\left(\begin{array}{ll} \smallskip E^+_b \\ E^-_b \end{array} \right)
\simeq m  \left( \begin{array}{ll}
\smallskip ~~\,{\rm sgn}\!\left[\cos(Vl / 2)\right]
\sin(Vl/2) \\ -\,{\rm sgn}\!\left[\sin(Vl / 2)\right]
\cos(Vl/2)   \end{array} \right).
\label{E_P_beta1}
\end{equation}

\begin{figure}[htb]
\begin{centering}
\includegraphics[width=0.99\textwidth]{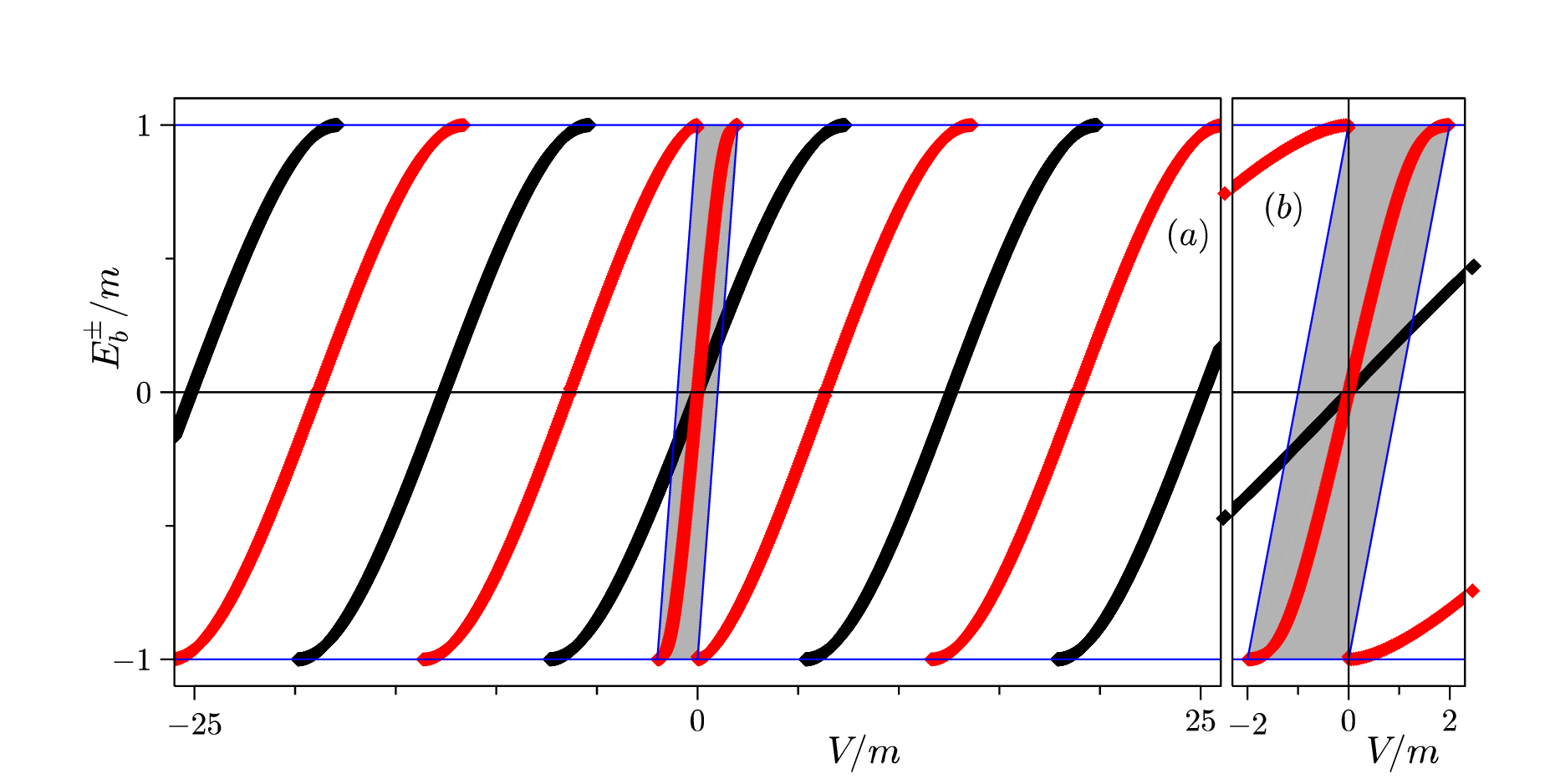}
\caption{Bound state energies $E_b^+$ (black lines) and $E_b^-$ (red lines) as
functions of strength $V \in \R$
for the line from pencil $P_1$  with $\alpha_1 =\alpha_2 =\alpha_3 =1$, 
$l=0.5\,m^{-1}$. Right panel:  Energies in region $|E-V|<m$. In shadowed area, 
 $k$ is imaginary.
 }
\label{fig4}
\end{centering}
\end{figure}

Notice that the lines of  $P_1$ satisfying the condition
 $\alpha_1 +\alpha_3 =2$ fall into  the ${\cal A}$-plane [see equation (\ref{rel_A})],
while  the lines of $P_2$ with $\alpha_1 =\alpha_3$ appear in the ${\cal B}$-plane
 [see equation  (\ref{rel_B})]. The other values of  $\alpha_1$ and $\alpha_3$
 correspond to the  potentials located outside  the flat band planes ${\cal A}$
 and ${\cal B}$. In the particular case of   $P_1$ with $\alpha_1 =\alpha_3=1$
 (the potential of type I), we have $\beta=1$ and, as a result, equations (\ref{bs_eqs}) reduce to
 \begin{equation}
 \!\!\!\!\!\!\!\!\!\!\!\!\!\!\!\!\!\!\!\!\!\!\!\!\!
 {\kappa \over k} \left(1 -{V \over E}\right) =
\left\{ \begin{array}{ll} \smallskip - \cot(kl/2) & \mbox{for}~ E=E^+, \\
~~\tan(kl/2) & \mbox{for}~E=E^- ,  \end{array} \right.\quad k=\sqrt{(E-V)^2-m^2}\,.
 \label{bs_typeI_eqs}
 \end{equation}
As follows from the form of these equations,  their solutions exist
 on the whole $V$-axis, where $k$ is real
(in the region $|E-V| >m$) and imaginary (in the region $|E-V| <m$), including
the lines $|E-V| =m$. For fixed $l$, these solutions are depicted
in figure~\ref{fig4}.
In the region $|E-V| <m$ ($k$ is imaginary), one of the solutions (for $E^+_b$) 
connects (on both the lines $|E-V|=m$) the pieces of the solution with $k>0$, while
the other solution for $E^-_b$
(completely located in the region $|E-V| <m$) appears as an additional branch
in the almost periodic series of the solutions displayed on the whole $V$-axis.

\subsection{Spectra with an asymptotically double bound states}

In the previous subsection, we have established that the sufficient condition
 for the periodicity of the bound state spectrum in the limit as $|V| \to \infty $
 is $\beta >0$, where $\beta$ is given by relation (\ref{beta}) and
 the coefficient  $\alpha_2 $ in both the pencils $P_1$ and $P_2$ is non-zero
 ($\alpha_2 =1$). Let us assume now that the parameter $\beta$ is negative. Then,
 setting ${\rm i}\sqrt{-\beta}$ in asymptotic equations
 (\ref{bs_P_eqs}) instead of $\sqrt{\beta}$,
  we arrive at the following two monotonic solutions  for large $V$:
\begin{equation}
\!\!\!\!\!\!\!\!\!\!\!\!\!\!\!\!\!\!\!\!\!\!\!\!\!\!\!\!\!\!\!\!\!\!\!\!\!\!\!\!\!\!
\left(\begin{array}{ll} \smallskip  E^+_b \\ E^-_b \end{array} \right)
\simeq {\rm sgn}(V)\,  m\left( \begin{array}{ll}
\smallskip \left[ 1 - \beta \coth^2(\sqrt{-\beta}\,Vl/2)\right]^{-1/2} \\
 \left[ 1- \beta\tanh^2(\sqrt{-\beta}\,Vl/2)\right]^{-1/2}  \end{array} \right)\!
 \to {\rm sgn}(V) {m \over \sqrt{1-\beta}}\,.
\label{E_D}
\end{equation}

In the particular case of the pencil $P_2 $ with $\alpha_1 =\alpha_3 =-1$ 
($\beta =-1$)
and $\alpha_2 =1$, according to (\ref{kP2}), we have $k = \sqrt{E^2 -V^2}$, so that
the explicit form of equations (\ref{bs_eqs}) in the cone region $|E| >|V|$
(where $k$ is real) becomes
\begin{equation}
\!\!\!\!\!\!\!\!\!\!\!\!\!\!\!\!\!
{\rm sgn}(E-V) {\kappa \over E}\sqrt{E-V \over E+V} =
 \left\{  \begin{array}{ll} \smallskip
 -\cot\!\left( \sqrt{E^2 -V^2}\, l/ 2\right) & \mbox{for}~~ E=E^+ , \\
 ~~\,\tan\!\left( \sqrt{E^2 -V^2}\,l/ 2\right) & \mbox{for}~~ E=E^- .
  \end{array} \right.
\label{bs_D_k_real_eqs}
\end{equation}

Beyond the cone, $k$ is imaginary and instead of equations (\ref{bs_D_k_real_eqs}),
we have
\begin{equation}
\!\!\!\!\!\!\!\!\!\!\!\!\!\!\!\!\!
{\rm sgn}(V-E) {\kappa \over E}\sqrt{V-E \over V+E} =
 \left\{  \begin{array}{ll} \smallskip
 -\coth\!\left( \sqrt{V^2-E^2 }\, l/ 2\right) & \mbox{for}~~ E=E^+ , \\
 ~~\,\tanh\!\left( \sqrt{V^2-E^2}\,l/ 2\right) & \mbox{for}~~ E=E^- .
  \end{array} \right.
\label{bs_D_k_imaginary_eqs}
\end{equation}

\begin{figure}[ht]
\begin{centering}
\includegraphics[width=0.99\textwidth]{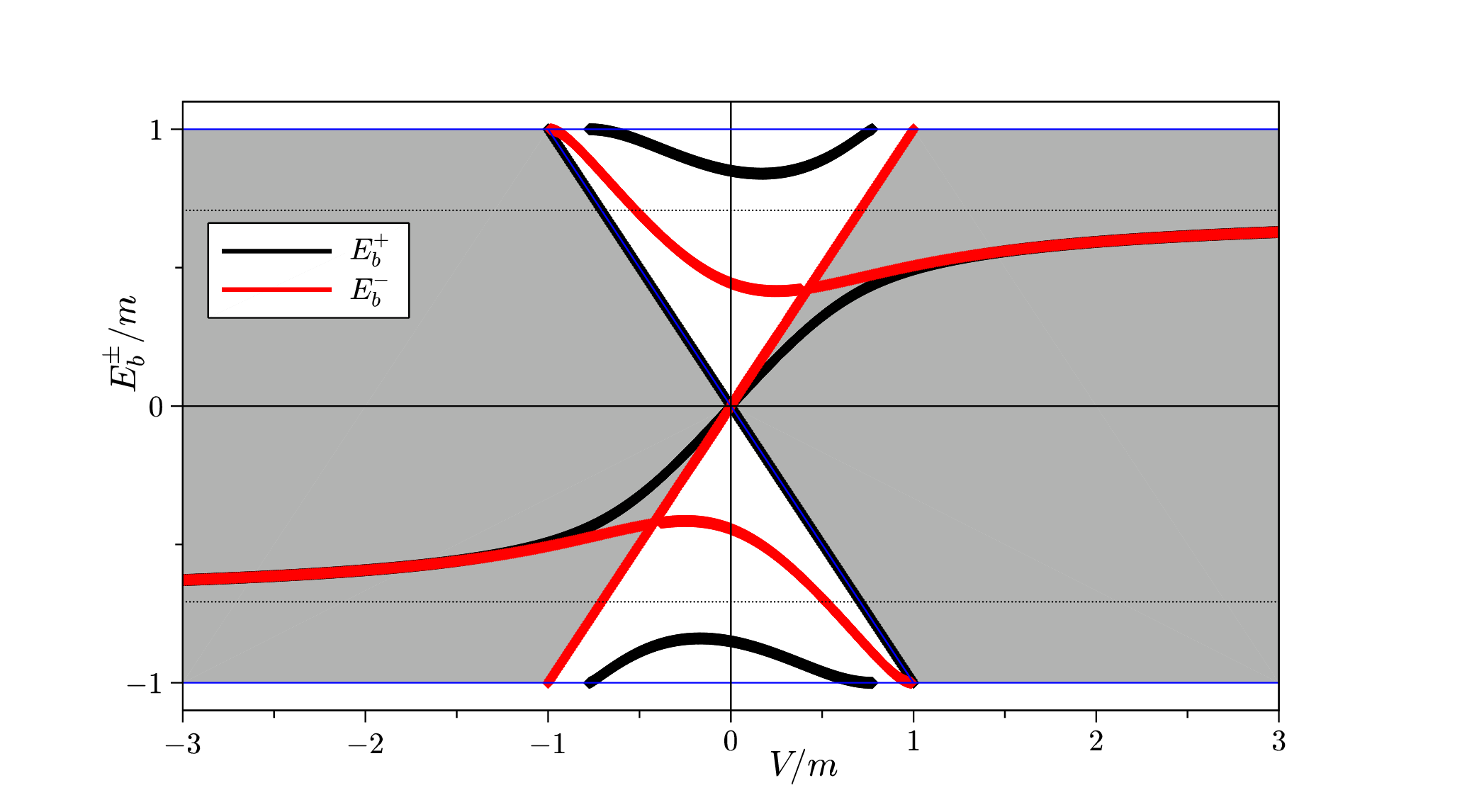}
\caption{Bound state energies $E_b^\pm$ as  functions of strength $V \in \R$
for the line from pencil $P_2$ with $\alpha_1 =\alpha_3 =- 1$ and $\alpha_2 =1$,
$l = 5\, m^{-1}$. In shadowed area, $k$ is imaginary.
}
\label{fig5}
\end{centering}
\end{figure}

 The cone boundary $|E|=|V|$ separates the regions with $k$ real and imaginary ($k=0$)
 and on this set  there are two simple solutions of this equation for $V \in (-m,m)$: $E_b^\pm =\mp V$.
The solutions of equations (\ref{bs_D_k_real_eqs}) and (\ref{bs_D_k_imaginary_eqs}) are depicted in figure~\ref{fig5}.
One can specify the $E^-_b$ solution as a ground state and the  $E^+_b$ solution 
as an excited state for $V<0$, while for $V>0$ their roles are reversed.

 \subsection{Spectra consisting of an infinite number of bound states}

Consider now the pencils $P_1$ and $P_2$, in which
  $\alpha_1 + \alpha_3 = 0$ and $\alpha_2 =1$.   Then, setting in this subsection
    $\alpha_1 \equiv \alpha =- \alpha_3$, for the pencil $P_1$,
  expressions (\ref{gamma}) and (\ref{kP1})  become
\begin{equation}
  \gamma = {\kappa \over k} \left( 1 - {V \over E}\right), \quad
   k=  \sqrt{ \left[ E^2 -(\alpha V +m )^2 \right] \left( 1 - {V \over E}\right) }\,.
   \label{k_gamma_H}
\end{equation}
For the pencil $P_2$, in this expression for $k$, it is sufficient to set formally
 $m=0$.  

 Assume first that $\alpha \neq 0$. Then, using for large $V$  expressions
 (\ref{k_gamma_H}), one can represent asymptotically equations (\ref{bs_eqs}) in the form
 \begin{equation}
 {\kappa V \over kE} \sim \left\{ \begin{array}{ll} \smallskip ~~\, \cot(kl/2)
 & \mbox{for}~E=E^+, \\ -\tan(kl/2) & \mbox{for}~E=E^-, \end{array} \right.
 \quad k\sim |\alpha|\sqrt{V^3 \over E}\,.
 \label{bs_eqs_H}
 \end{equation}
  Since $V/k \to 0$ as  $|V| \to \infty$, and taking  into  account that $|E| < m$,
 from the right-hand sides of  equations (\ref{bs_eqs_H}), we obtain the following
 approximate solutions for the bound state energies for large $V$:
\begin{equation}
E_b^\pm = E_n \simeq \left( {\alpha l \over n \pi}\right)^{\!\!2} \!V^3,
\quad |V| < \left( {n\pi \over \alpha l}\right)^{\!2/3}m^{1/3},
\quad n=1,\, 2, \ldots ,
\label{E_n_H}
\end{equation}
 where odd $n$'s stand for $E^+_b$ and even $n$'s for $E^-_b$.  Here, with
 increasing the $n$th level, the energies $E_n$ are successively cutting
 at the thresholds $E= \pm m$.

\begin{figure}[ht]
\begin{centering}
\includegraphics[width=0.99\textwidth]{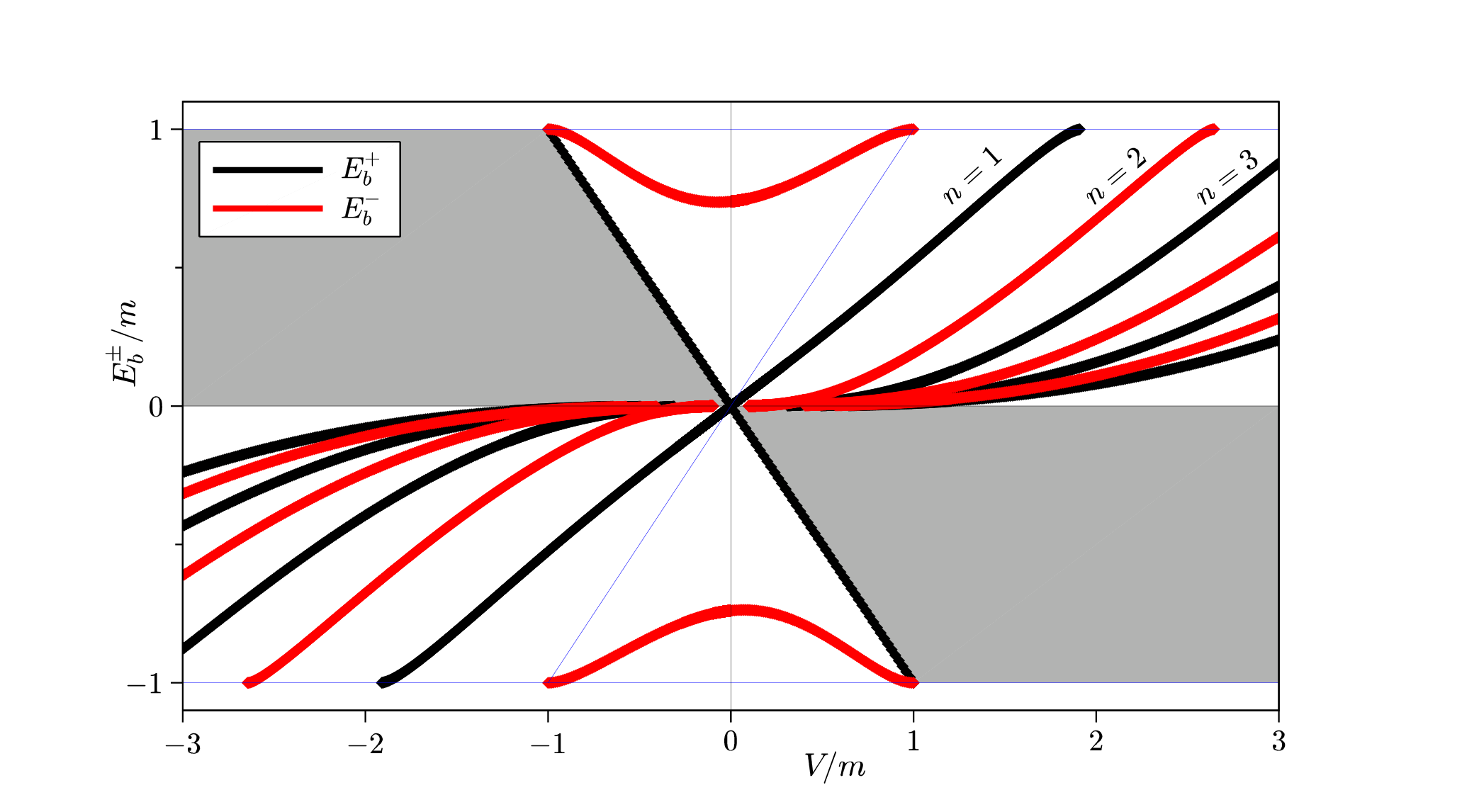}
\caption{Bound state energies $E_b^\pm$ as  functions of strength $V \in \R$
for the line from pencil $P_2$ with $\alpha_1 = \alpha_2 =- \alpha_3=1 $, 
$l=2\,m^{-1}$. In shadowed area,  $k$ is imaginary.
 }
\label{fig6}
\end{centering}
\end{figure}

 For the illustration of the behavior of the bound state energies
 on the whole $V$-axis, let us consider the line in the pencil $P_2$,
 for which $\alpha_1 = \alpha_2 =- \alpha_3=1 $. Then, equations (\ref{bs_eqs})
 take the explicit form as follows
 \begin{equation}
{\rm sgn}\!\left(1 -{V \over E}\right) {\kappa \over \sqrt{E(E+V)}} =
 \left\{  \begin{array}{ll} \smallskip
 -\cot(k l/ 2) & \mbox{for}~~ E=E^+ , \\
 ~~\,\tan(kl/ 2) & \mbox{for}~~ E=E^- ,
  \end{array} \right.
\label{bs_H_eqs}
\end{equation}
with  $k= \sqrt{(E^2 -V^2)(1 - V / E)}$ [instead of $k$ in
(\ref{k_gamma_H})]. The series of exact solutions of
  equations (\ref{bs_H_eqs}) is depicted in figure~\ref{fig6}, one of which
  is  simple: $E^+_b = -V$. In this figure,
 the region where $k$ is real consists of the cone $|E| > |V|$
 plus the two strips ($0<E< m$,  $0<V< \infty$, $E<V$) and
  ($- m <E< 0$, $-\infty <V< 0$,  $E>V$). In the region consisting of the
  two strips ($0<E< m$,  $-\infty <V< 0$, $E + V<0$) and
  ($-m <E< 0$,  $0<V< \infty$, $E+V>0$), $k$ is imaginary. Setting
$k= {\rm i}\sqrt{(V^2 -E^2)(1 - V / E)}$  into equations (\ref{bs_H_eqs})
 and taking into account that $1 -V/E >0$ in these strips, we conclude that
 the left- and right-hand sides of equations have opposite signs. Therefore
  there are no solutions in the region  where $k$ is imaginary.
Since at the thresholds $E=\pm m$, we have $\kappa =0$, the cutoff   of the
bound state energies $E_{b}$ is determined by the solutions of the equations
 $\cot(kl/2)=0$ and $\tan(kl/2)=0$ or correspondingly by the explicit equations
 $(V+m)^2(1-V/m)=(n\pi/l)^2$ for $V\le -m$ and
 $(V-m)^2(1+V/m)=(n\pi/l)^2$ for $V\ge m$.

\begin{figure}[ht]
\begin{centering}
\includegraphics[width=0.75\textwidth]{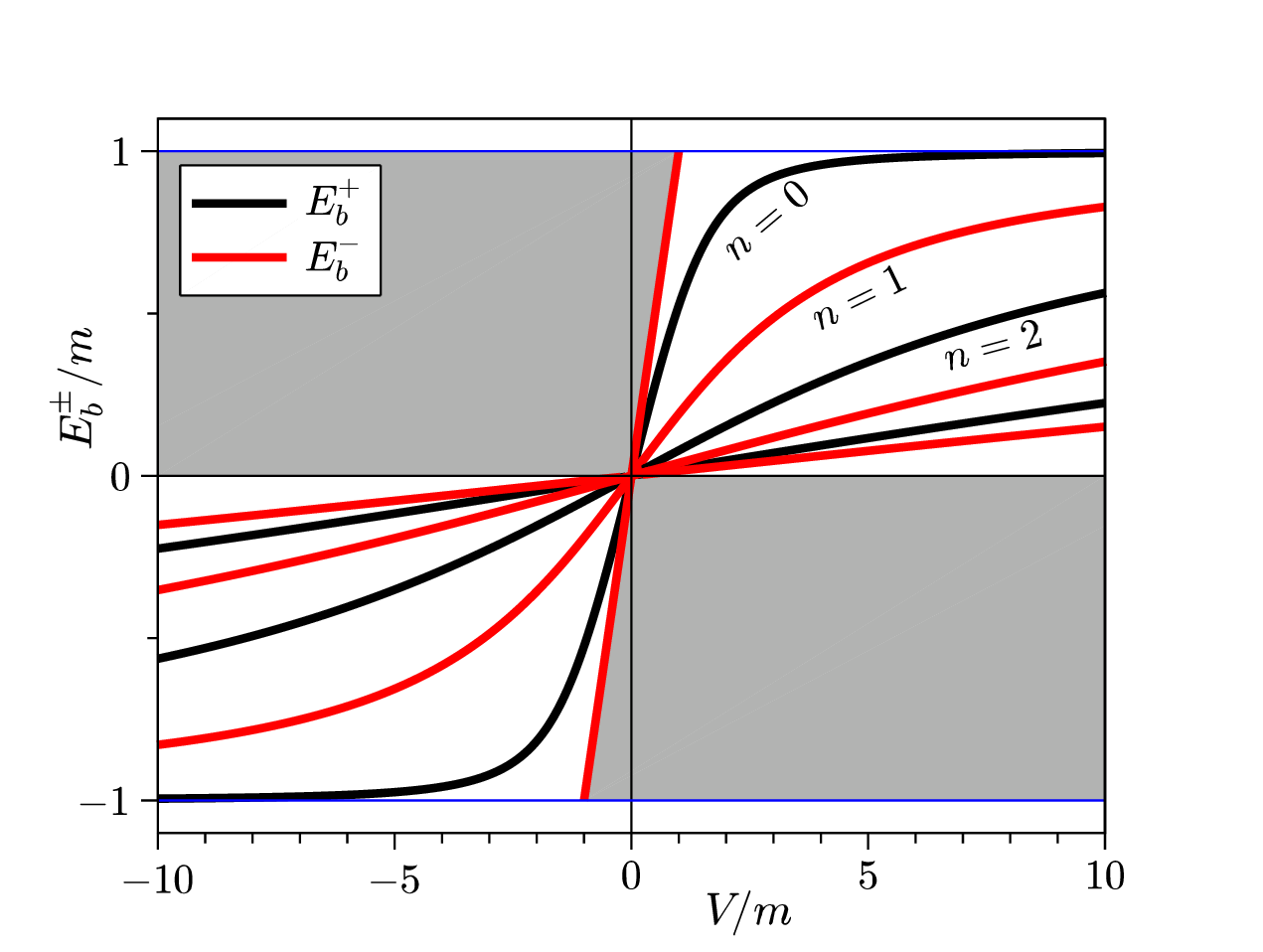}
\caption{Bound state energies $E_b^\pm$ as  functions of strength $V \in \R$
for the line from pencil $P_1$ with $\alpha_1 = \alpha_3 = 0$ and $\alpha_2=1 $,
$l=2\,m^{-1}$. In shadowed area,  $k$ is imaginary.
 }
\label{fig7}
\end{centering}
\end{figure}

A similar spectrum consisting of an infinite number of bound state energies,
 but without cutoffs at the thresholds $E = \pm m$, takes also place
  in the particular case $\alpha =0$, where we are dealing with
  the potential of type II ($V_{11}=V_{33} =0$,  $V_{22} =V$), studied in
\cite{Zhang2022JPB}. More precisely, this is the line from the pencil $P_1$
with $\alpha_1 =\alpha_3 =0$ and $\alpha_2 =1$, which does not belong to either
the ${\cal A}$- or  ${\cal B}$-planes.  Thus, owing to (\ref{k}) and (\ref{gamma}),
 we have
\begin{equation}
 k= \kappa \sqrt{{V \over E}-1}\,, \quad \gamma =  - \sqrt{{V \over E}-1}\,,
 \label{k_gamma_H1}
\end{equation}
 so that  equations (\ref{bs_eqs}) can be rewritten in the explicit form as  follows
\begin{equation}
 \sqrt{{V \over E}-1} =\left\{  \begin{array}{ll} \smallskip
 ~~\, \cot\!\left[(\kappa l/ 2)\sqrt{V/ E-1}\,\right] & \mbox{for}~~ E=E^+ , \\
 -\tan\!\left[(\kappa l/ 2)\sqrt{V/ E-1}\,\right]& \mbox{for}~~
 E=E^- .  \end{array} \right.
\label{bs_H1_eqs}
\end{equation}
Here, $k$ is real in the two strips: ($0<E< m$, $0<V< \infty$ and $E<V$) and 
($-m <E<0$, $-\infty <V<0$ and $E>V$).
In the case if $k$ is imaginary, the left- and right-hand sides of equations (\ref{bs_H1_eqs})  have opposite signs,
therefore there are no solutions with imaginary $k$. The solution $E^- =V $
that corresponds to $k=0$, splits
the regions of real and imaginary $k$'s. This means that the sign of
the bound state energy $E_b^+$
must coincide with the sign of the strength $V$, i.e., the bound state energies $E_b^\pm$
must be both positive if $V>0$, and negative if $V < 0$. The solution of equations
(\ref{bs_H1_eqs}) on the whole $V$-axis is depicted in figure~\ref{fig7}, where 
it is  shown that the $E_{b}^+$- and $E_{b}^-$-levels   alternate.
Here, as follows from the form of  equations
  (\ref{bs_H1_eqs}), for each $V$, there exists  an infinite number of energy levels.

  For large $V$,  the approximate solution of equations (\ref{bs_H1_eqs}),
  which coincides with formula (33) in \cite{Zhang2022JPB}, reads
\begin{equation}
\!\!\!\!\!\!\!\!\!\!\!\!\!\!\!\!
E_b^\pm =E_{n} \simeq {\rm sgn}(V)\sqrt{{(n\pi/l)^4 \over 4V^2} +m^2}
-{(n\pi/l)^2 \over 2V} <m, \quad n=1,\,2\, \ldots,
\label{E_n_II}
\end{equation}
where even $n$'s stand for $E_{b}^+$ and odd $n$'s for $E_{b}^-$.
There is also a solution that corresponds to $n=0$:
\begin{equation}
E_b^+=E_{0} \simeq {\rm sgn}(V)\frac{m}{\sqrt{1+ (2/Vl)^2}},
\label{E_0_II}
\end{equation}
which faster approaches the threshold values $E=\pm m$ as $|V|\to\infty$.
If $V\to 0$, the energy of all the levels is proportional to the potential strength
($E_n\propto  V)$.

\subsection{Bound states with a successive detachment from the thresholds}

In this subsection, we describe the spectrum of bound states, the number of which
is finite for each fixed strength $V$, however, this number increases 
with growth of $V$
because of a successive detachment of new bound states from the thresholds
$E = \pm m$. To this end, let us consider both the pencils $P_1$ and $P_2$
with $\alpha_2 =0$. Notice that the lines
with  $\alpha_1 = -\alpha_3$ in the pencil $P_1$ fall into the ${\cal A}$-plane
($V_{11} + V_{33} =2V_{22}=0$), while the lines with $\alpha_1 =\alpha_3$ in
the pencil $P_2$ appear in the ${\cal B}$-plane ($V_1 = V_3)$.

In general, if  $\alpha_2 =0$, but $\alpha_1 \neq 0$ and  $\alpha_3 \neq 0$,
equations (\ref{bs_eqs}) for both the pencils $P_1$ and $P_2$  can be solved
exactly on the whole $V$-axis. The solutions exist only if $k>0$ because
for imaginary $k$, both the sides of equations (\ref{bs_eqs}) have opposite signs,
 including the limit $k \to 0$.

\begin{figure}[htb]
\begin{centering}
\includegraphics[width=0.9\textwidth]{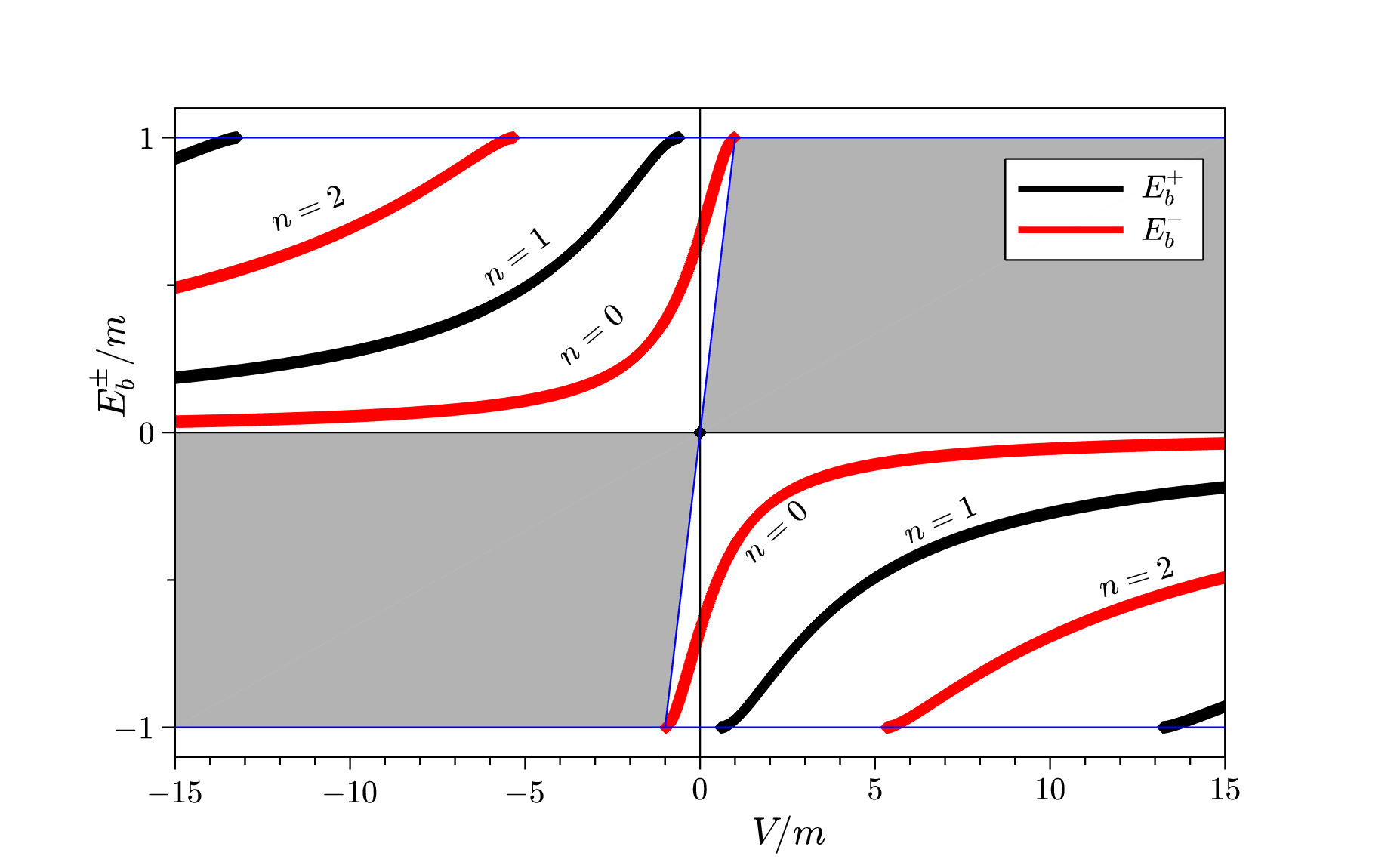}
\caption{Bound state energies $E_{b}^\pm$ as  functions of
 strength $V \in \R$ for the line from pencil $P_1$ with $\alpha_1 =\alpha_3 =1$
 and $\alpha_2 =0$,  $l=2.5\,m^{-1}$.  In shadowed area,  $k$ is imaginary.
 }
\label{fig8}
\end{centering}
\end{figure}

Analytically,  one can investigate
the asymptotic behavior of the bound state spectrum if $V$ is sufficiently large.
Thus, for both the pencils $P_1$ and $P_2$,  we have
$k \sim \sqrt{-\beta EV}$ with $\beta$  given by (\ref{beta}), so that
asymptotically for large $V$, equations (\ref{bs_eqs}) become
\begin{equation}
\!\!\!\!\!\!\!\!\!\!\!\!\!\!\!\!\!\!\!\!\!\!
 {\kappa \over \sqrt{-\beta EV}} \sim \left\{  \begin{array}{ll} \smallskip
 -\cot\!\left(\sqrt{-\beta EV}\,l/ 2\right)      & \mbox{for}~~ E =E^+ \,, \\
  ~\,\, \tan\!\left(\sqrt{-\beta EV}\,l/ 2\right)  & \mbox{for}~~ E =E^- \,.
  \end{array} \right.
 \label{bs_W_eqs}
\end{equation}
Solving these asymptotic equations and using that $|E|<m$, we get
\begin{equation}
\!\!\!\!\!\!\!\!\!\!\!\!\!\!\!\!\!\!\!\!
E_b^\pm = E_n \simeq - { (n\pi/ l )^2 \over \beta V}\,,\quad
{(n\pi/l)^2 \over |\beta | m} < |V| < \infty , \quad  n=1,\,2, \ldots ,
\label{E_n_W}
\end{equation}
where odd $n$'s stand for $E^+_b$ and even $n$'s for $E^-_b$. One more solution,
\begin{equation}
E_b^- = E_0 \simeq  -{\rm sgn}(\beta V) {m \over \sqrt{1 + (\beta V l/2)^2} }
\end{equation}
that corresponds to $n=0$,  is obtained studying the limit as $EV \to 0$.
These solutions are illustrated by figure~\ref{fig8}, where an exact solution
to equations (\ref{bs_eqs}) on the
whole $V$-axis is represented for the particular case $\beta =1$. In this case,
in equations (\ref{bs_eqs}), we
have $k= \sqrt{E(E-V)}$ and $\gamma = \kappa/k$.

One can consider another configuration of the coefficients $\alpha_j$ in the pencils
$P_1$ and $P_2$, namely $\alpha_2 \neq 0$ but $\alpha_3 =0$. Let us consider
the configuration $\alpha_1 \equiv \alpha > 0$, $\alpha_2 =1$ and $\alpha_3 =0$
in the pencil $P_1$. Then, for large $V$, we have $k \sim \sqrt{-\alpha (m+E)V}$ and
therefore equations (\ref{bs_eqs}) become
\begin{equation}
\!\!\!\!\!\!\!\!\!\!\!\!\!\!\!\!\!\!\!
 {  \kappa  \over  E}\sqrt{- {V \over \alpha (m+E)}} \sim \left\{ \begin{array}{ll}
 \smallskip  - \cot\!\left(\!\sqrt{-\alpha (m+E)V}\,l/2\right) & \mbox{for}~~ E =E^+ \,, \\
 ~~\, \tan\!\left(\!\sqrt{-\alpha (m+E)V}\,l/2\right)  & \mbox{for}~~ E =E^- \,.
 \end{array} \right.
 \label{bs_W1_eqs}
\end{equation}
Solving these asymptotic equations for large $V$ and using that $|E |< m$, we get
the following approximate solution:
\begin{equation}
\!\!\!\!\!\!\!\!\!\!\!\!\!\!\!\!\!\!\!\!
E_b^\pm = E_n \simeq -\left[m +  { (n\pi/ l )^2 \over \alpha V}\right],\quad
- \infty < V < - {(n\pi/l)^2 \over 2\alpha m}\,, \quad n=1,\,2, \ldots ,
\label{E_n_W1}
\end{equation}
where even $n$'s stand for $E^+_b$ and odd $n$'s for $E^-_b$. Except for this solution, there exists also a solution
(assigned by the number $n=0$) that approaches the thresholds $E=\pm m$  more rapidly.
For $k >0$, only the first equation (\ref{bs_W1_eqs}) admits a solution in the limit as $E \to -m$, which must be negative.
This solution is valid only for $V <0$ and it  coincides with that given by formula  (\ref{E_0_II}). On the other hand,
for positive  $V$, $k$ is imaginary and, as a result, in the limit as $E \to m$ and
$V \to \infty$, both equations
(\ref{bs_W1_eqs}) have also solutions, which  coincide in the limit as $V \to \infty$.
Thus, on the whole $V$-axis, the
$n=0$ bound state energy solution approximately reads
\begin{equation}
E_b^+ = E_0 \simeq m \left\{ \begin{array}{ll} \smallskip
- \left(1+ 4/V^2l^2\right)^{\!-1/2} & \mbox{for}~V<0, \\
~~\, \left(1+ 2\alpha m /V \right)^{\!-1/2} & \mbox{for}~V>0. \end{array} \right.
\label{E_0_W1}
\end{equation}

\begin{figure}[htb]
\begin{centering}
\includegraphics[width=0.9\textwidth]{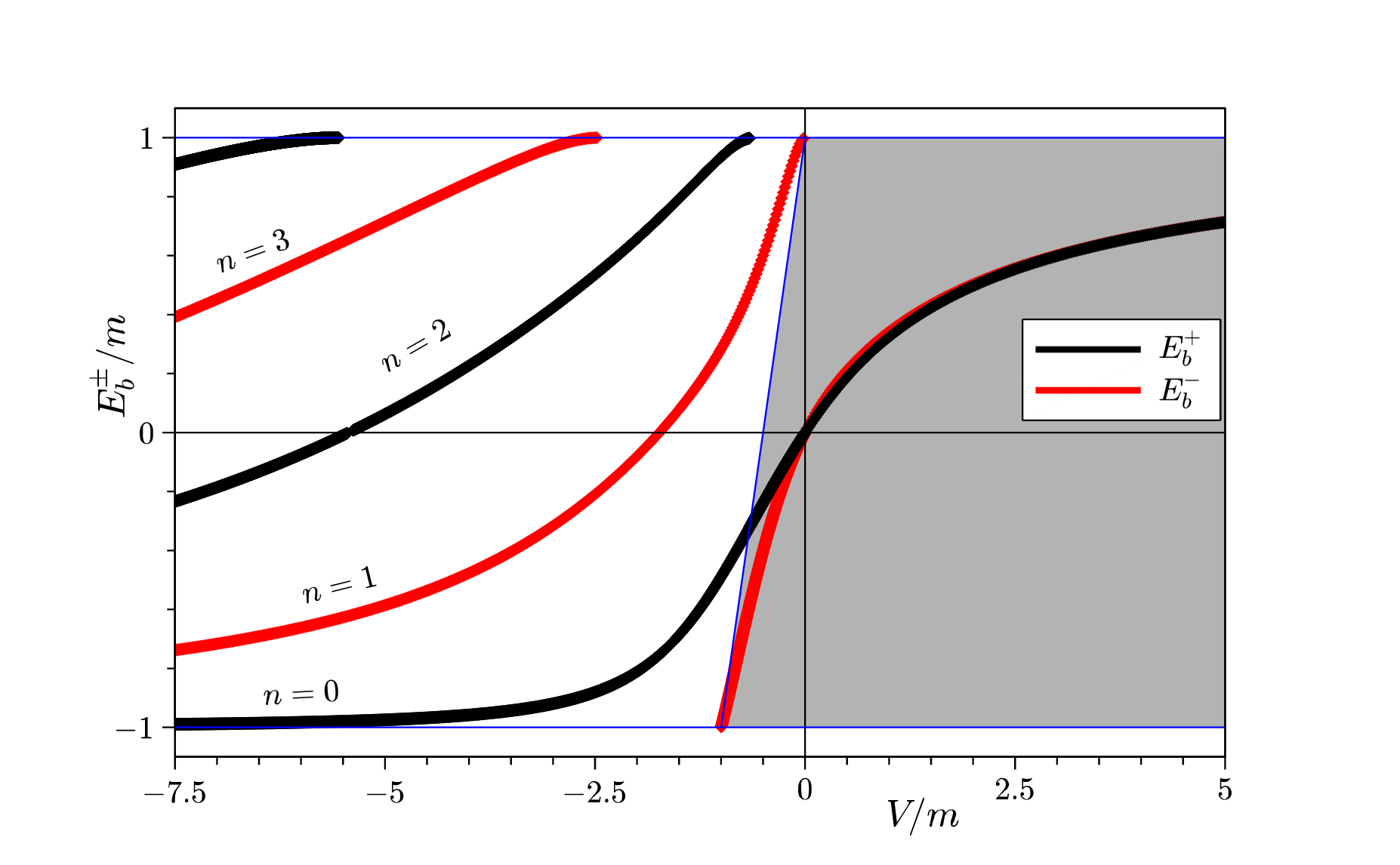}
\caption{Bound state energies $E_b^\pm$ as functions of strength $V \in \R$
for the line from pencil $P_1$ with $\alpha_1=2$, $\alpha_2 =1$ and
$\alpha_3 =0$, $l=2\,m^{-1}$. In shadowed area,  $k$ is imaginary.
 }
\label{fig9}
\end{centering}
\end{figure}

Consider the particular case of  the line from the pencil $P_1$ with
$ \alpha_1 =2$, $\alpha_2 =1$ and $\alpha_3 =0$.
 This line falls into the ${\cal A}$-plane and,
  as follows from representation  (\ref{kP1}),  $ k = \sqrt{(m+E)(E-2V-m)}$\,,
so that equations (\ref{bs_eqs}) with this $k$ become
\begin{equation}
 {  \kappa \over k }\left(1 -{V \over  E}\right) =\left\{ \smallskip 
 \begin{array}{ll}
 -\cot(kl/ 2)      & \mbox{for}~~ E =E^+ \,, \\ ~~ \tan(kl/ 2) &
  \mbox{for}~~ E =E^- \,. \end{array} \right.
 \label{bs_W_alpha2_eqs}
\end{equation}
 Solving these equations, we obtain the bound state spectrum on
 the whole $V$-axis, which is depicted in figure~\ref{fig9}.
%
%
 In the region where $E > 2V+m$ and $E \in (-m,m)$, we have
$k>0$ and, as a result, along the negative half-axis $V$, the successive
detachment of bound state energies occurs. For imaginary $k$, there are
two solutions, which are displayed for positive $V$. In the limit as $V \to \infty$,
these solutions merge to a single bound state energy.

Thus, we have examined some types of bound state spectra, in which
the energy levels $E=E_b^\pm$ crucially depend on the strength components
in the $(V_{11},\,V_{22},\,V_{33})$-space. The set of admissible vectors in this
space has been restricted by the two pencils of straight lines $P_1$ and $P_2$
 with the vertices at
the points $(0,\,0,\,0)$ and $(-m,\, 0,\,m)$. As defined by equations (\ref{rep1})
and (\ref{rep2}), both the pencils are parametrized by the strength parameter $V$
and the coefficients $\alpha_j \in \R$, $j=1,\,2,\,3$. Therefore, for a given set
of these coefficients,  it is possible to describe the bound state energies
 as functions of the  parameter $V$.
According to the asymptotic behavior of the energy levels for large values of $V$,
from the whole variety of spectra, we have singled out at least
 four characteristic species, referred in the following to as $P$, $D$, $H$ and $W$
 types:
\begin{itemize}
\item[(i)] The spectra of type $P$ are described by the two-valued  almost periodic
levels $E= E_b^+$ and $E= E_b^-$ as illustrated by figure~\ref{fig4}. This type
is realized on the set
\begin{equation}
\!\!\!\!\!\!\!\!\!\!\!\!\!\!\!\!\!\!\!\!\!\!\!\!\!\!\!\!
A_P := \left\{\alpha_j~\mbox{in}~ P_1~\mbox{and}~P_2 ~\Big|~
\alpha_j \neq 0, \, j=1,\,2,\,3,
~{\alpha_1 \alpha_3 \over \alpha_1 + \alpha_3 }>0\right\}.
\label{A_P}
\end{equation}
The spectra of this type are periodic in the limit as $|V| \to \infty$.
\item[(ii)]
The spectra of type $D$ are described by the
 double-valued levels $E = E_b^+$ and $E=E_b^-$ with a monotonic behavior
for $|V| >m$. In the limit as $|V| \to \infty$, the double levels
merge to single levels. This type is materialized on the set
\begin{equation}
\!\!\!\!\!\!\!\!\!\!\!\!\!\!\!\!\!\!\!\!\!\!\!\!\!\!\!\!
A_D:= \left\{\alpha_j~\mbox{in}~ P_1~\mbox{and}~P_2 ~\Big|~
\alpha_j \neq 0, \, j=1,\,2,\,3,
~{\alpha_1 \alpha_3 \over \alpha_1 + \alpha_3 }<0 \right\}
\label{A_D}
\end{equation}
and  illustrated by figure~\ref{fig5}.
\item[(iii)]
The spectra  of type $H$, consisting of an infinite number of energy levels
that obey the law $E_b^\pm = E_n \propto n^{-2}$, $n \in \N$, resemble
 the hydrogen atom spectrum. One of the spectra of type $H$ includes
 the levels with successive
cutoffs at the thresholds $E =\pm m$ as illustrated by figure~\ref{fig6},
so that at a given $V$, the number of levels is finite but it increases to
infinity as $|V| \to \infty$. This spectrum is realized on the set
\begin{equation}
A_{H,1}:= \left\{\alpha_j~\mbox{in}~ P_1~\mbox{and}~P_2 ~\Big|~
\alpha_1  =- \alpha_3 \neq 0,~ \alpha_2 \neq 0 \right\}.
\label{A_{H,1}}
\end{equation}
The other $H$-spectrum, shown in figure~\ref{fig7}, is materialized on the set
\begin{equation}
A_{H,2}:= \left\{\alpha_j~\mbox{in}~ P_1~\Big|~
\alpha_1  =\alpha_3 = 0,~ \alpha_2 \neq 0 \right\}.
\label{A_{H,2}}
\end{equation}
Unlike the previous spectrum, the number of levels is infinite at a given $V$
for this spectrum.
\item[(iv)]
The spectra of type $W$, consisting of the energy levels
$E_b^\pm = E_n \propto n^{2}$, $n \in \N$,
with  their successive detachment from the thresholds $E=\pm m$,
 resemble the spectrum of a potential well with its increasing depth but fixed width.
 One of these spectra, illustrated by figure~\ref{fig8}, is realized   on the set
 \begin{equation}
 \!\!\!\!\!\!\!\!\!\!\!\!\!\!\!\!\!\!\!\!\!\!\!\!\!\!\!\!
A_{W,1}:= \left\{\alpha_j~\mbox{in}~ P_1~\mbox{and}~P_2 ~\Big|~
\alpha_1  \neq 0,~\alpha_3 \neq 0,~\alpha_1 +\alpha_3 \neq 0,~
 \alpha_2 = 0 \right\}.
\label{A_{W,1}}
\end{equation}
   The other $W$-spectrum, shown in  figure~\ref{fig9}, is materialized
   on the set
 \begin{equation}
A_{W,2}:= \left\{\alpha_j~\mbox{in}~ P_1 ~\Big|~
\alpha_1  > 0,~\alpha_2 \neq 0,~\alpha_3 =0 \right\}.
\label{A_{W,2}}
\end{equation}
 \end{itemize}

\section{Point interactions realized from  rectangular potentials}

  One-center point interactions can be obtained as
the rectangular potentials of type (\ref{V(x)}) are squeezed to a point.
 Particularly,  based on equations (\ref{bs_eqs}), one can materialize one-point
 interactions with finite bound state energies $E_b$ for various types of
  the potential $V(x)$ in the squeezing limit as $l \to 0 $. To accomplish this
  limit properly, first we have to derive the asymptotic behavior of $k$, $\gamma$
  and $\eta$ for small $l$ [according to expressions (\ref{k}), (\ref{gamma})
  and (\ref{eta})] and then to find  the $l \to 0$ limit of equations (\ref{bs_eqs})
  for bound state energies and finally connection matrix (\ref{L_matrix}).

   One of these point interactions is the $\delta$-limit of the rectangular potentials
\begin{equation}
\!\!\!\!\!\!\!\!\!\!\!\!\!\!\!\!\!\!\!\!\!
V_{jj}(x) =  {g \over l}\left\{ \begin{array}{ll} 1 &
\mbox{for}~~ x_1 \le x \le x_2,\\
                               0 & \mbox{otherwise}           \end{array}
 \right. \to g\delta(x)\quad \mbox{as}\,\, l\to 0,\,\, ~~~g \in \R,
 \label{delta(x)}
\end{equation}
where $g$ is a dimensionless strength constant of the $\delta$-potential.
This a particular case of the regularization of a delta function.
More generally, for the approximation of the potential $g\delta(x)$ by regular
functions in the one-dimensional Dirac equation, a
scaled sequence $l^{-1}h(x/l)$ with $\int_{-\infty}^\infty h(\xi)d\xi=g$
has been applied in paper \cite{Tusek2020}. It should be emphasized that,
as proven in this work, the realized 
point interactions do not depend on the shape of 
the function $h(\xi)$.  In our case, as shown below
for some examples of the rectangular potential $V(x)$, the $l^{-1}$-approximation
is valid only for ground bound states and it does not `cover' excited states.
To describe properly the excited states in a one-point approximation,
another type of squeezing is used
below, namely with the strength parameter $V \sim g/l^2m$ as $l \to 0$.
 In the following, we refer  this type of squeezing  to as a `$l^{-2}$-limit'.
Another type of squeezing to be used
is the asymptotic representation $V \sim  g(m/l^2)^{1/3}$, referred to as
a `$l^{-2/3}$-limit'.
Thus, using below the formulas for $k$,  $\gamma$ and $\eta$, we will calculate  
these squeezing limits of equations   (\ref{bs_eqs}) for several configurations
 of the potentials $V_{jj}(x)$, $j=1,\,2,\,3$.

We perform the  $l \to 0$ limit at the origin of the $(x_1,\,x_2)$-plane,
 setting  first $x_1 \to -0$ and then $x_2 \to +0$ as one of the ways of
 approaching the origin. In this case, $a \to 0$ but the repeated limit
 of the ratio $l/a$ is finite:
$\lim_{x_2 \to +0}\lim_{x_1 \to -0}(l/a) =2$. Using then the second relation
(\ref{C_1_C_2}), wave functions (\ref{psi+_out}) and (\ref{psi-_out}) are
transformed to the form that involves only one arbitrary constant $C_1$: 
\begin{equation}
\!\!\!\!\!\!\!\!\!\!\!\!\!
\begin{array}{ll} \smallskip
\!\!\!\!\!\!\!\!\!\!\!\!\!\!\!\!\!\!\!\!\!\!\!\!
 \psi^{+}(x)=C_1  \left\{ \begin{array}{ll} \smallskip
  {\rm col}\!\left(\rho^{-1}, \, \sqrt{2}\,, \, \rho
 \right)\!{\rm e}^{\kappa x}, & -\infty < x < 0, \\
  {\rm col}\!\left(-\rho^{-1}, \,  \sqrt{2}\,, \,- \rho
 \right)\!{\rm e}^{-\kappa x}, & ~0 < x < \infty ,
 \end{array} \right.~~\mbox{for}~ E=E^+, \\
 \!\!\!\!\!\!\!\!\!\!\!\!\!\!\!\!\!\!\!\!\!\!\!\!
 \psi^{-}(x)= C_1  \left\{ \begin{array}{ll} \smallskip
  {\rm col}\!\left(\rho^{-1}, \, \sqrt{2}\,, \, \rho
 \right)\!{\rm e}^{\kappa x}, & -\infty < x < 0, \\
  {\rm col}\!\left(\rho^{-1}, \, - \sqrt{2}\,, \, \rho
 \right)\!{\rm e}^{-\kappa x}, & ~0 < x < \infty ,
  \end{array} \right. ~~~~\mbox{for}~ E=E^-.
  \end{array}
 \label{psi_out_0}
\end{equation}
 Here, $E = E_b^\pm$ are the $l \to 0$ limit values of the solutions to
equations (\ref{bs_eqs}). Explicitly, using that $\rho = (m-E)/\kappa$ and
$\rho^{-1} =(m +E)/\kappa$, the two-sided (at $x = \pm 0$)  boundary conditions
for bound states can be represented in the following form:
\begin{equation}
\begin{array}{ll} \smallskip
\psi^+(\pm 0)= C_1 {\rm col}\left( \mp (m +E^+_b)/\kappa^+_b, \, \sqrt{2}, \,
\mp (m -E^+_b)/\kappa^+_b \right), \\
\psi^-(\pm 0)= C_1 {\rm col}\left(  (m +E^-_b)/\kappa^-_b, \, \mp \sqrt{2}, \,
 (m -E^-_b)/\kappa^-_b \right), \end{array}
 \label{bcs}
 \end{equation}
where $\kappa^\pm_b := \kappa(E_b^\pm)$. Then  $\Lambda$-matrix
(\ref{L_matrix}) connects the squeezed boundary conditions of
 the components $(\psi^\pm_1-\psi^\pm_3)(x)$ and $\psi^\pm_2(x)$:
\begin{equation}
\!\!\!\!\!\!\!\!\!\!\!\!\!\!\!\!\!\!\!\!\!\!\!\!\!\!\!\!\!\!\!\!\!\!
\!\!\!\!\!\!\!\!\!\!
\left(\!\!\!\begin{array}{cc} \smallskip
 (\psi_1^+ -\psi_3^+)(\pm 0) \\ \psi_2^+(\pm 0) \end{array} \!\! \right)=
     C_1 \left(\!\!  \begin{array}{ll} \smallskip \mp 2E^+_b /\kappa^+_b \\
~~~ \sqrt{2}          \end{array}\!\!\!\right),~
\left(\!\!\!\begin{array}{cc} \smallskip
 (\psi_1^- -\psi_3^-)(\pm 0) \\ \psi_2^- (\pm 0) \end{array} \!\!\right)=
  C_1    \left( \!\! \begin{array}{ll} \smallskip  2E^-_b /\kappa^-_b \\
\, \mp \sqrt{2}        \end{array}\!\!\!\right).~~
 \label{psi+-0}
\end{equation}

Finally, we note that an infinite number of bound states  also exists
 for the potential of type III ($V_{11}  = V \in \R\setminus \{0\}$
 and  $ V_{22} = V_{33} \equiv 0$), as proven in \cite{Zhang2022JPB}.
However, in the limit as $l \to 0$,
 for $k$ [see relation (\ref{k})] and $\gamma$ we have the
limits: $k \to  \sqrt{2E (m+E)}$ and $\gamma \to \sqrt{(m-E)/2E}$. Since both these
expressions are finite and $kl \to 0$,  equations (\ref{bs_eqs})  have no solutions,
i.e., there are no bound states for the potential of type III in the squeezing limit.

\subsection{The $\delta$-limit }

{\it Type $P$}:
The $\delta$-limit of the bound states of type $P$ is obtained immediately by
 replacing the product  $Vl$ in solutions (\ref{E_P}) and (\ref{E_P_beta1}) with
 the strength $g$. In formula (\ref{E_P_beta1}), the solution can be combined in
 the form of  the two-valued periodic (increasing) discontinuous function
 $E = m {\cal E}(g)$, where ${\cal E}(g)$ is  constructed from the piece
\begin{equation}
 \varepsilon(g) = \left( \begin{array}{ll} \varepsilon^+(g) \\ \varepsilon^-(g) \end{array}
 \right) : = \left( \begin{array}{ll} ~~ \sin(g/2) & \mbox{for}~~  0 \le g < \pi \\
              - \cos(g/2) & \mbox{for}~~0< g \le \pi  \end{array} \right)
 \label{varepsilon}
\end{equation}
that repeats itself  forward and backward along the $g$-axis.
The period of the function ${\cal E}(g)$ is  $\pi$, so that ${\cal E}(g +\pi)={\cal E}(g)$,
$g \in \R$. Furthermore, since  $kl \to \sqrt{\beta}|g|$ and
$\eta \to - {\rm sgn}(g)\sqrt{2/\beta}$
[see equation (\ref{eta})], in the limit as $l \to 0$,
$\Lambda$-matrix (\ref{L_matrix}) reduces to the form
\begin{equation}
 \Lambda  =\left( \begin{array}{lc}   \smallskip
  ~~\cos\!\left(\sqrt{\beta}g\right)~~~~~~
 -\!\sqrt{2/\beta} \sin\!\left(\sqrt{\beta}g\right) \\
        \!\!\sqrt{\beta/2}\sin\!\left(\sqrt{\beta}g\right)~~~~~~~
        \cos\!\left(\sqrt{\beta}g\right)   \end{array} \right)
        \label{L_matrix_P}
\end{equation}
where $\beta >0$. Using expressions (\ref{E_P}) for the squeezed energies $E_b^\pm$,
one can check that matrix (\ref{L_matrix_P})
 connects the two-sided boundary values of components (\ref{psi+-0})
at $x =\pm 0$.

{\it Type $D$}: Similarly, replacing ${\rm sgn}(V)$ and $Vl$ in bound state energy
(\ref{E_D}) with ${\rm sgn}(g)$ and $g$, respectively,
we obtain the $\delta$-limit of the squeezed energies $E_b^\pm$.
In this case, the connection matrix is described by the same formula (\ref{L_matrix_P})
where $\beta <0$. In a similar way, using equations (\ref{E_D}), one can check
that the  boundary values (\ref{psi+-0}) are connected by matrix (\ref{L_matrix_P})
with  $\beta < 0$.

{\it Type $H$}:
For realizing a point interaction that corresponds to the ground state,
we use the $\delta$-limit defined by (\ref{delta(x)}),
setting $V \sim g/l$ in equations (\ref{bs_H_eqs}). Hence,
 we have $\sqrt{V/E -1} \sim \sqrt{g/E\,l}$ and only the first of equations
(\ref{bs_H1_eqs}) admits a finite solution in the limit as $l \to 0$.
 Explicitly,  this equation
reduces to  $\kappa^+_b/E^+_b =2/g$, having the  solution for  the bound state energy:
\begin{equation}
 E^+_b  =E_{0} =  { m g  \over \sqrt{4 +g^2}}
\label{E_H_ground_energy}
\end{equation}
that  coincides exactly with formula (13) in \cite{Zhang2022JPB}.

Furthermore, in the limit as $l \to 0$, we have $kl \to 0$ and,
according to definition (\ref{eta}),
$\eta \sin(kl) \to -\sqrt{2}\,g$. Therefore,  matrix (\ref{L_matrix}), connecting
the two-sided boundary conditions (\ref{psi+-0})
for the ground state energy $E_0$\,, becomes
\begin{equation}
\!\!\!\!\!\!\!\!\!\!\!\!\!\!\!\!
\Lambda = \Lambda_0  =\left( \begin{array}{lc}          1~~~~-\sqrt{2}\,g \\
        0~~~~~~~~ ~1   \end{array} \right)=        \left( \begin{array}{lc}
        1~~~~-2\sqrt{2}\,E_0/\sqrt{m^2 -E_0^2} \\
        0~~~~~~~~~~~~ ~1   \end{array} \right).
\label{L_matrix_H_ground}
\end{equation}

{\it Type $W$}: Consider  the realization of the $\delta$-limit for the pencils
$P_1$ and $P_2$ with $\alpha_1 \neq 0$, $\alpha_2 = 0$ and $\alpha_3 \neq 0$.
Setting $V \sim g/l$ in expressions (\ref{kP1}) and (\ref{kP2}), we find that
 $k \sim \sqrt{-\beta gE/l}$ where $\beta gE <0$. Using this asymptotic
representation in equations (\ref{bs_W_eqs}), one can see that only the equation for $E^-$
admits a solution in the $l \to 0$ limit. Indeed, in this limit, the second equation
(\ref{bs_W_eqs}) reduces to $\kappa/E^-=-\beta g/2$, resulting to the ground state energy
\begin{equation}
 E^-_b = E_0 = -{\rm sgn}(\beta g){m \over \sqrt{1+\beta^2 g^2/4}}\,.
 \label{bs_W_energy}
\end{equation}
Furthermore, $\eta = \sqrt{2}\, E / k \sim \sqrt{2} \,E / \sqrt{-\beta gE/l}$
and therefore $\eta \sin(kl) \to 0 $,
while $-\eta^{-1}\sin(kl) \to \beta g/\sqrt{2}$\,.
Thus, the connection matrix becomes
\begin{equation}
\!\!\!\!\!\!\!\!\!\!\!\!\!\!\!\!
 \Lambda =\Lambda_0  =\left( \begin{array}{cc}        ~~ ~ 1~~~~~~~~~\, 0 \\
       \beta g/ \sqrt{2}~~~~~1    \end{array} \right) =\left( \begin{array}{cc}
        ~ ~ ~~~~~~~1~~~~~~~~~~~~~~~~~~~~\, 0 \\
        -\sqrt{2(m^2 -E_0^2)}/E_0 ~~~~~1    \end{array} \right).
        \label{L_matrix_W}
\end{equation}

For the other configuration with $\alpha_1 \equiv \alpha > 0$, $\alpha_2 =1$ and
$\alpha_3 =0$, using the representation $V \sim g/l$,
we get $k \sim \sqrt{-\alpha (m+E)g/l}$\,. Using this relation in equations (\ref{bs_W1_eqs}), we find that only the first equation
for $E^+$ admits a finite limit, i.e., $2E^+ =g\kappa$, which can be solved explicitly.
Taking  into  account that $gE>0$, the solution coincides with expression (\ref{E_H_ground_energy}). Furthermore, we have
the limit $\eta \sin(kl) \to -\sqrt{2}\,g$,
resulting in the same connection matrix (\ref{L_matrix_H_ground}).

\subsection{The $l^{-2/3}$-limit}

{\it Type $H$}:
For realizing the point interactions that describe the series of bound states
(\ref{E_n_H}), we use the asymptotic representation $V \sim g(m/l^2)^{1/3}$ 
with a dimensionless strength  $g \in \R\setminus \{0\}$. 
Then, $E_n \to (\alpha/n\pi)^2g^3m$ defined on the intervals
 $0< |g| < (n\pi/\alpha)^{2/3}$ and $kl \to |\alpha| \sqrt{g^3m/E} = n\pi$,
 $n =1, \, 2, \ldots $, with odd $n$'s for $E^+_b$ and even $n$'s for $E_b^-$.
 Due to these relations as well as equations (\ref{bs_eqs_H}),  using that
 $\sin^2(kl/2) =1$ if $E_n=E^+_b$  and $\cos^2(kl/2)=1$ if $E_n=E^-_b$, 
 one can use the  following representation:
  \begin{eqnarray}
 \!\!\!\!\!\!\!\!\!\!\!\!\!\!\!\!\!\!\!\!\!\!\!\!\!
 \sin\!\left( \!|\alpha| \sqrt{g^3m \over E}\, \right)  &=& 2\left\{ 
  \begin{array}{ll}
 \smallskip  \sin^2\!\left[(|\alpha|/2) \sqrt{g^3m/E}\right]
 \cot\!\left[(|\alpha|/2) \sqrt{g^3m/E}\right]  \\
  \cos^2\!\left[(|\alpha|/2) \sqrt{g^3m/E}\right]
  \tan\!\left[ (|\alpha|/2) \sqrt{g^3m/E}\right]
 \end{array} \right. \nonumber \\
  &\sim & 2(-1)^{n+1} \,{2 \kappa V\over kE }\, ,
\end{eqnarray}
 where $V/k \to 0$ as $l \to 0$. On the other hand, $\eta \sim - \sqrt{2}\,V/k$ and,
 as a result, $\eta \sin(kl) \to 0$ and  $-\eta^{-1}\sin(kl) \to
 (-1)^{n+1}\sqrt{2(m^2 - E^2)}/E$ as $l \to 0$. Since
 $\cos(kl) = (-1)^n$, connection matrix (\ref{L_matrix}) becomes
 \begin{equation}
 \!\!\!\!\!\!\!\!\!\!\!\!\!\!\!\!\!\!\!\!\!\!\!\!\!\!\!\!\!\!\!\!\!\!\!\!\!\!\!\!\!\!
 \Lambda = \Lambda_n = (-1)^n \left(\!\! \begin{array}{cc} \smallskip
 ~~~~~~~~~~~1  ~~ ~~~~~~~~~~~~~~~~~0 \\ -\sqrt{2(m^2 -E^2_n)}/E_n ~~~~~~1
 \end{array}\!\!\right), ~ ~E_n = \left({\alpha \over n\pi}\right)^2 g^3m,
 ~~n \in \N.
 \label{L_matrix_H}
\end{equation}

\subsection{The $l^{-2}$-limit}

{\it Type $H$}:
For realizing the point interactions that describe the excited bound states with
energies (\ref{E_n_II}), we use the asymptotic representation $V \sim g/l^2m$ with a
dimensionless strength  $g \in \R\setminus \{0\}$. Then equations (\ref{bs_H1_eqs})
asymptotically become
\begin{equation}
 \sqrt{mE \over g}\,l \,\sim \,\left\{  \begin{array}{ll} \smallskip
 ~~\, \tan\!\left[(\kappa/ 2)\sqrt{g/ mE}\,\right] & \mbox{for}~~ E=E^+ , \\
 -\cot\!\left[(\kappa / 2)\sqrt{g / mE}\,\right]& \mbox{for}~~
 E=E^- .  \end{array} \right.
\label{bsII_eqs_asymp}
\end{equation}
In the limit as $l \to 0$, both these asymptotic relations lead to the series of
equations  $kl \to \kappa \sqrt{g / mE} = n\pi$, $n=1, 2, \ldots $,
where $E=E^+_b$ stands for even $n$'s and $E=E^-_b$ for odd $n$'s.
The solution of these equations with respect to $E \in (-m,m)$ reads
\begin{equation}
\!\!\!\!\!\!\!\!\!\!\!\!
  E_{b}^\pm = E_n =
 {n^2 \pi^2 m \over 2g } \left( \sqrt{1+ {4 g^2 \over n^4\pi^4}} -1\right)
 \simeq {gm \over n^2\pi^2}\,,~~~n \in \N,~~
 \label{bsII_excited_energies}
\end{equation}
supporting the $1/n^2$ law only for the excited states.  Note that $E_{n} \in (0,m)$
if $g >0$ and $E_{n} \in (-m,0)$ if $g <0$.

Furthermore, from (\ref{eta}) we have $\eta   \sim
-(g/ \kappa l) \sqrt{2E/ mg}$\,. On the other hand, using representation
(\ref{bsII_eqs_asymp}), we obtain
\begin{eqnarray}
 \sin\!\left(\!\kappa \sqrt{g \over mE}\,\right) &=& 2\left\{  \begin{array}{ll}
 \smallskip
 \tan\!\left[(\kappa / 2)\sqrt{g / mE}\,\right]
 \cos^2\!\left[(\kappa / 2)\sqrt{g / mE}\,\right] \\
 \sin^2\!\left[(\kappa / 2)\sqrt{g / mE}\,\right]
 \cot\!\left[(\kappa/ 2)\sqrt{g / mE}\,\right]  \end{array} \right. \nonumber \\
 & \sim & 2(-1)^n \,\sqrt{mE \over g}\,l ,
\end{eqnarray}
where even $n$'s correspond to $E^+_b$ and odd $n$'s to $E^-_b$. As a result,
in the limit as $l \to 0$, we obtain
$\eta \sin(kl) \to -2\sqrt{2}\, (-1)^n E_n/\sqrt{m^2 -E_n^2}$. Since $\cos(kl) \to (-1)^n$, taking for account matrix
(\ref{L_matrix_H_ground}),  the connection matrix for all the bound states reads as follows
\begin{equation}
\!\!\!\!\!\!\!\!\!\!\!\!\!\!\!\!\!\!\!\!\!\!\!\!\!\!
 \Lambda = \Lambda_n = (-1)^n \left(\!\! \begin{array}{cc} \smallskip
 1  ~~~~-2\sqrt{2}\, E_n/\sqrt{m^2 -E^2_n}  \\ 0 ~~~~~~~~~~~~~~~~~~~~~~1~~~~~~~~~~
 \end{array}\!\!\right),~~~~n \in \N\cup \{0\},
 \label{L_matrixII_n}
\end{equation}
where $E_n =E_{b}^+$ for even $n$ and $E_n =E_{b}^-$ for odd $n$. Here, the ground state energy $E_0$ is given by expression
(\ref{E_H_ground_energy}) and excited state energies by equations (\ref{bsII_excited_energies}).
 These energies are arranged as $m > |E_0| > |E_1|> \ldots > |E_n| >\ldots$ for
 all $g \in \R \setminus \{0\}$.  Notice that
in matrix (\ref{L_matrixII_n}), for $n=1,2, \ldots$, we have $E_n / \sqrt{m^2 -E_n^2}
 \simeq  g / n^2\pi^2$\,.

 {\it Type $W$}:  To implement point interactions that describe the excited bound states
 for the pencils $P_1$ and $P_2$ with $\alpha_1,\, \alpha_3 \neq 0$ and 
 $\alpha_2 =0$, we substitute the asymptotic representation
 $V \sim g/l^2m$ into equations (\ref{bs_W_eqs}). As a result,  these equations 
 become
\begin{equation}
  {\kappa \over \sqrt{-\beta gE/m}} \,l \, \sim \, \left\{  \begin{array}{ll} \smallskip  -\cot(\sqrt{-\beta gE/m}/ 2)      & \mbox{for}~~ E =E^+ \,, \\ ~~\,
 \tan(\sqrt{-\beta gE/m}/ 2) & \mbox{for}~~ E =E^- \,.
 \end{array} \right.
 \label{bsIV_eqs_asymp}
\end{equation}
In the limit as $l \to 0$, from these equations we obtain the solution
\begin{equation}
 E_b^\pm = E_n = - {n^2 \pi^2 m \over \beta g}\,,~~~n \in \N,
 \label{sol_IV}
\end{equation}
where odd $n$'s stand for $E_b^+$ and even $n$'s for $E_b^-$. Using further 
asymptotic representation (\ref{bsIV_eqs_asymp}), one can write
\begin{eqnarray}
 \sin \sqrt{-\beta g E \over  m} &=& 2\left\{  \begin{array}{ll} \smallskip
 \sin^2\!\left[\sqrt{-\beta gE/ m} / 2\right]
 \cot\!\left[\sqrt{-\beta gE/ m}  / 2\right]  \\
 \tan\!\left[ \sqrt{-\beta gE/ m} / 2\right]
 \cos^2\!\left[\sqrt{-\beta gE/  m} / 2\right]
 \end{array} \right. \nonumber \\
  &\sim & 2(-1)^n \,{ \kappa \over \sqrt{-\beta gE/m}}\,l ,
\end{eqnarray}
where odd $n$'s correspond to $E^+_b$ and even $n$'s to $E^-_b$.
Therefore, in the limit as $l \to 0$, we have $ \eta^{-1} \sin(kl) \to
(-1)^n   \sqrt{2(m^2 -E_n^2)}\,/E_n\,.$
Since $\cos(kl) \to (-1)^n$, taking for account matrix (\ref{L_matrix_W}),
 the connection matrix for all the bound states reads as follows
\begin{equation}
\!\!\!\!\!\!\!\!\!\!\!\!\!\!\!\!\!\!\!\!\!\!\!\!\!
 \Lambda = \Lambda_n = (-1)^n \left(\!\! \begin{array}{cc} \smallskip
~~~~~~ ~~~1 ~~~~~~~~~~~~~~~~~~~~~~0  \\ - \sqrt{2(m^2-E_n^2)}\,/E_n~~~~~~~1~
 \end{array}\!\!\right),~~~~n \in \N\cup \{0\},
 \label{matrixIV_n}
\end{equation}
where the bound state energies $E_n$ are given by expressions
 (\ref{bs_W_energy}) and (\ref{sol_IV}) with
$E_n =E_{b}^+$ for odd $n$ and $E_n =E_{b}^-$ for even $n$. Here
$ E_0 $ is a ground state energy and the energies $E_n$ with
$n=1,\,2, \ldots $ correspond to
excited states. These energies are arranged as
$m > |E_0| > |E_1|> \ldots > |E_n| >
\ldots$ for all $g \in \R \setminus \{0\}$.

Similarly, for  the case of the pencil $P_1$
with $\alpha_1 \equiv \alpha >0$, $\alpha_2=1$ and $\alpha_3 =0$,
owing to the relation $V \sim g/l^2m$, we have
$k \sim \sqrt{-\alpha(1 +E/m)g}/l$ with $g<0$.
Further, the asymptotic representation of equations (\ref{bs_W1_eqs}) reads
 \begin{equation}
 \!\!\!\!\!\!\!\!\!\!\!\!\!
  {E \sqrt{\alpha m(m+E)} \over \kappa \sqrt{-g}}\,l
 \,\sim \, \left\{  \begin{array}{ll} \smallskip
 -\tan \!\left[ \sqrt{-\alpha (1 +E/m)g} / 2 \right]
   & \mbox{for}~~ E =E^+ \,, \\ ~~\,
 \cot\!\left[ \sqrt{-\alpha (1 +E/m)g} / 2\right]
  & \mbox{for}~~ E =E^- \,.  \end{array} \right.
 \label{bs_W1_eqs_asymp}
\end{equation}
In the limit as $l \to 0$, from these equations, we obtain
$\sqrt{-\alpha (1 + E/m) g} =n\pi,~ n =1,\,2, \ldots $,
where even $n$'s stand for $E^+$ and odd $n$'s for $E^-$. Solving the last
 equation and taking for account that $|E|<m$, we get the series of excited
 bound states with the energies
\begin{equation}
\!\!\!\!\!\!\!\!\!\!\!\!\!\!\!\!\!\!\!\!\!\!\!\!\!\!
 E_b^\pm =E_n = - \left( 1+ {n^2 \pi^2 \over \alpha g}\right)\! m,
 \quad -\infty < g < - {n^2\pi^2 \over 2\alpha}\,,~~~n \in \N.
 \label{E_n_W1_energy}
\end{equation}
These energies are detached successively from the upper threshold $E=m$.
Using relations (\ref{E_H_ground_energy})
and (\ref{E_n_W1_energy}), one can prove that the bound state energies  are arranged
in the order  $E_0< E_1 < \ldots E_n < \ldots$.

 Using further asymptotic representation (\ref{bs_W1_eqs_asymp}), one can write
\begin{eqnarray}
\!\!\!\!\!\!\!\!\!\!\!\!\!\!\!\!\!\!\!\!\!\!\!\!\!\!\!\!\!\!\!\!
 \sin \!\sqrt{-\alpha (1 +E/m)g}  &=& 2\left\{  \begin{array}{ll} \smallskip
 \cos^2[\sqrt{-\alpha (1 +E/m)g}/ 2]\, \tan[\sqrt{-\alpha (1 +E/m)g}/ 2]  \\
 \sin^2[\sqrt{-\alpha (1 +E/m)g}/ 2] \,  \cot[\sqrt{-\alpha (1 + E/  m)g}/2 ]
 \end{array} \right.  \nonumber \\
 &\sim & 2(-1)^{n+1} \,{ E\sqrt{\alpha m(m+E)} \over \kappa \sqrt{- g}}\,l ,
\label{bs_W1_eqs_asymp_n}
\end{eqnarray}
where even $n$'s correspond to $E^+_b$ and odd $n$'s to $E^-_b$.
Next, in the limit as $l \to 0$, we have
 $ \eta \sin(kl) \to  (-1)^{n+1} 2\sqrt{2}\, E_n/ \sqrt{m^2 -E_n^2}$\,.
Since $\cos(kl) \to \cos\sqrt{-\alpha (1+E/m)g} = (-1)^n$, taking for account matrix
(\ref{L_matrix_H_ground}),  the connection matrix for all the bound states is the same
as for the potential of type II given by matrix (\ref{L_matrixII_n}).
 Here, the energies $E_n$ with $n=1,\,2, \ldots $ correspond to the excited states.

Thus, the equations derived above for the bound state energies
in the squeezing limit indicate that only those energy levels, which are
stretched on the $V$-axis to infinity, as illustrated by figures from \ref{fig4} to \ref{fig9}, admit a point approximation. The levels with a finite support, which
are shown in figures~\ref{fig4}--\ref{fig7} and \ref{fig9}, are not appropriate
for implementing point interactions. The energies $E_b^\pm$ in the squeezing limit
become functions of the dimensionless strength constant $g$.

The point interactions considered above are determined by the matrices
that connect the two-sided boundary conditions for a wave functions $\psi^\pm(x)$ at
the origin $x =\pm 0$. To implement these interactions, we have applied three
rates of squeezing as $l \to 0$. One of these is the $l^{-1}$-limit
resulting in the typical $\delta$-interaction. In this limit, for the spectra
of types $P$  and $D$, the connection matrix is given by (\ref{L_matrix_P}),
 where $\beta >0$ and $\beta <0$ correspond to perfectly periodic energies (\ref{E_P})
 and  to double-valued energies (\ref{E_D}), respectively, with ${\rm sgn}(V)$ 
 and $Vl$ substituted by ${\rm sgn}(g)$ and $g$.   For the spectra of types
$H$ and $W$, the $l^{-1}$-limit leads to the existence of ground states,
which are indicated in table~\ref{tab:table1} with $n=0$.

The $l^{-2/3}$- and $l^{-2}$-limits generate the countable sets
of point interactions that describe the excited states in the $H$- and
$W$-spectra. As indicated in table~\ref{tab:table1}, for these interactions,
there are two connection matrices
\begin{equation}
\!\!\!\!\!\!\!\!\!\!\!\!\!\!\!\!\!\!\!\!\!\!\!\!\!\!\!\!\!\!\!\!\!\!\!\!\!\!\!\!\!\!\!
\Lambda_n = \left(\begin{array}{cc} ~1 ~~~~0 \\ 2\chi_n ~~1 \end{array} \right)
~\mbox{and}~\left(\begin{array}{cc} 1 ~~~ 2/\chi_n \\ \!\!\!\!
 0 ~~~~~1~\end{array} \right), ~~ \chi_n := -{\sqrt{(m^2 -E_n^2)/2} \over E_n },
 ~~n \in \N\cup\{0\}. ~~~
\end{equation}
It should be noticed that the $l^{-2}$-limit has been applied in many publications
(see, e.g., \cite{zpi,Golovaty2009,Golovaty2013,ZZ2014,ZZ2021,Seba,
Griffiths,Christiansen}, a few to mention), mainly for regularizing
a potential in the form of the derivative of a delta function in the non-relativistic Schr\"{o}dinger equation.

\begin{table}
\caption{\label{tab:table1} Squeezed connection matrices and bound state
energies for point interactions of types $H$ and $W$ obtained in the $l \to 0$ limit
with three rates $l^{-1}$, $l^{-2}$ and $l^{-2/3}$. The energy levels, shown in
figures \ref{fig6}--\ref{fig9}, which admit the realization of
the point interactions, are  indicated in the last column. 
 }
\begin{tabular}{cccccc}
\hline
\hline
Connection &  $A$-sets & Types   & Squeezing &  Bound state & Energy \\
matrices $\Lambda_n $ & & of spectra & limits & energies   & levels in figures\\
\hline  \\
 $(-1)^n \left(\!\! \begin{array}{cc} ~ 1 ~~~~\,0 \\ 2\chi_n ~~1 \end{array} \! \right)$
& $A_{H,1}$ & $H$ & $l^{-2/3} $ & $E_n$ in (\ref{L_matrix_H}) & 6 ($n \in \N $) \\
    \par\smallskip
  & $A_{W,1}$ &$W$ & $l^{-1}$ & $E_0$ in (\ref{bs_W_energy}) & 8 ($n=0$) \\
  \par\smallskip
  & $A_{W,1}$ &$W$ & $l^{-2}$ & $E_n$ in (\ref{sol_IV}) & 8 ($n \in \N$) \\
 \hline \\
 $(-1)^n \left( \! \! \begin{array}{cc} 1 ~~~2/\chi_n \\ 0~~~ ~~1~~~
\end{array} \!\!\! \right)$ & $A_{H,2}$ & $H$ & $l^{-1} $ & $E_0$ in
(\ref{E_H_ground_energy}) & 7 ($n=0$) \\
    \par\smallskip
    &  $A_{H,2}$ & $H$ & $l^{-2} $ & $E_n$ in (\ref{bsII_excited_energies})
     & 7 ($n \in \N $)  \\
   \par\smallskip
  &  $A_{W,2}$ & $W$ & $l^{-1} $ & $E_0$ in (\ref{E_H_ground_energy})
   & 9 ($n =0 $)  \\
   \par\smallskip
  &  $A_{W,2}$ & $W$ & $l^{-2} $ & $E_n$ in (\ref{E_n_W1_energy})
  & 9 ($n \in \N $)  \\
\hline
\hline
\end{tabular}
\end{table}

\begin{figure}[htb]
\begin{centering}
\includegraphics[width=0.8\textwidth]{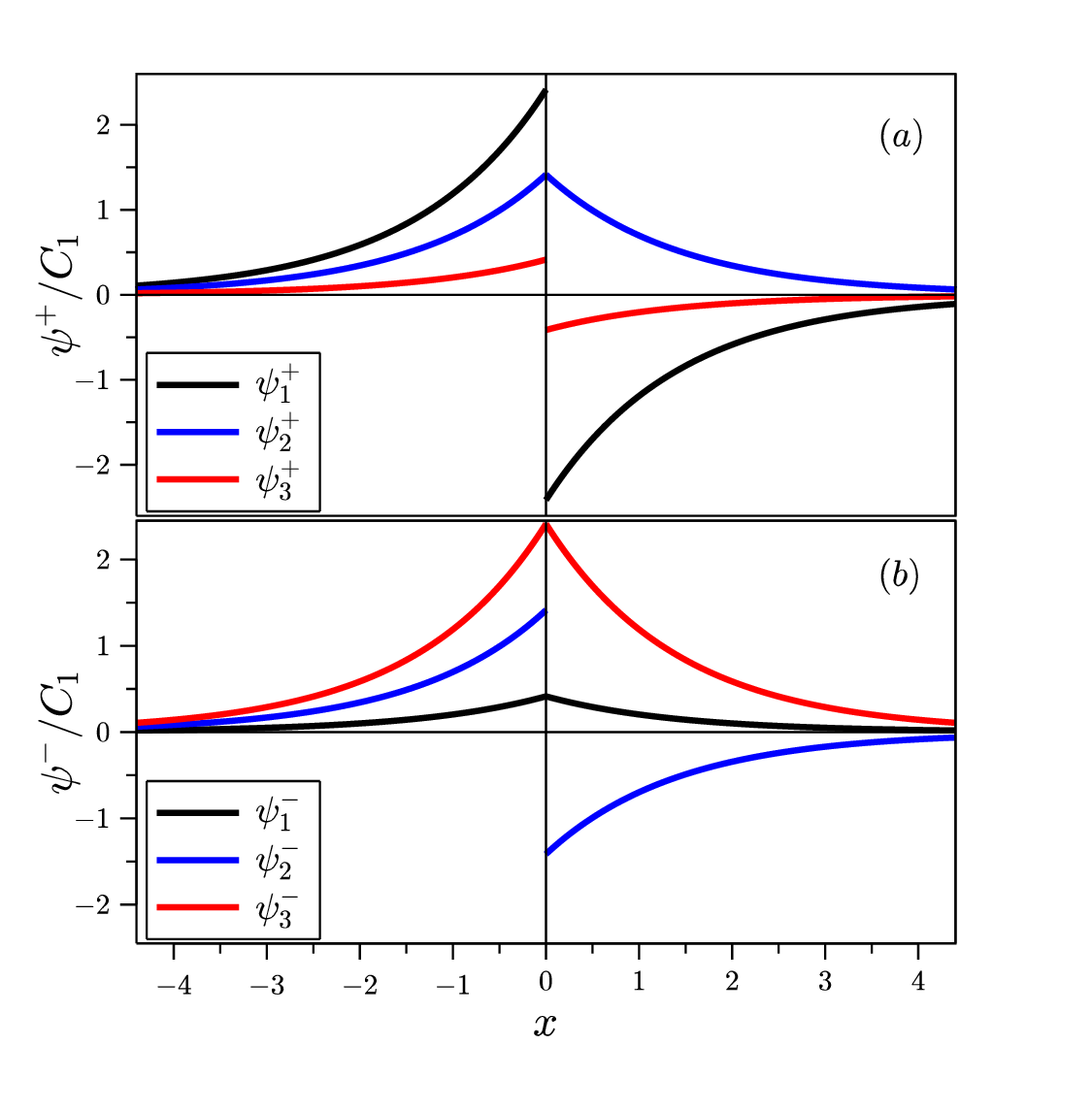}
\caption{Squeezed eigenfunctions $\psi^\pm(x)$ that correspond to  spectrum
of type $P$, which have been plotted according to formulas (\ref{psi_out_0}).
 Here,  the bound state energies $E_b^\pm$,
 obtained in the $l^{-1}$-limit, are  given by equations (\ref{varepsilon}):
 (a)~$E_b^+/m = \sin(g/2)$ and (b)~$E_b^-/m = - \cos(g/2)$ with strength $g=\pi/2$. }
\label{fig10}
\end{centering}
\end{figure}

\begin{figure}[htb]
\begin{centering}
\includegraphics[width=0.8\textwidth]{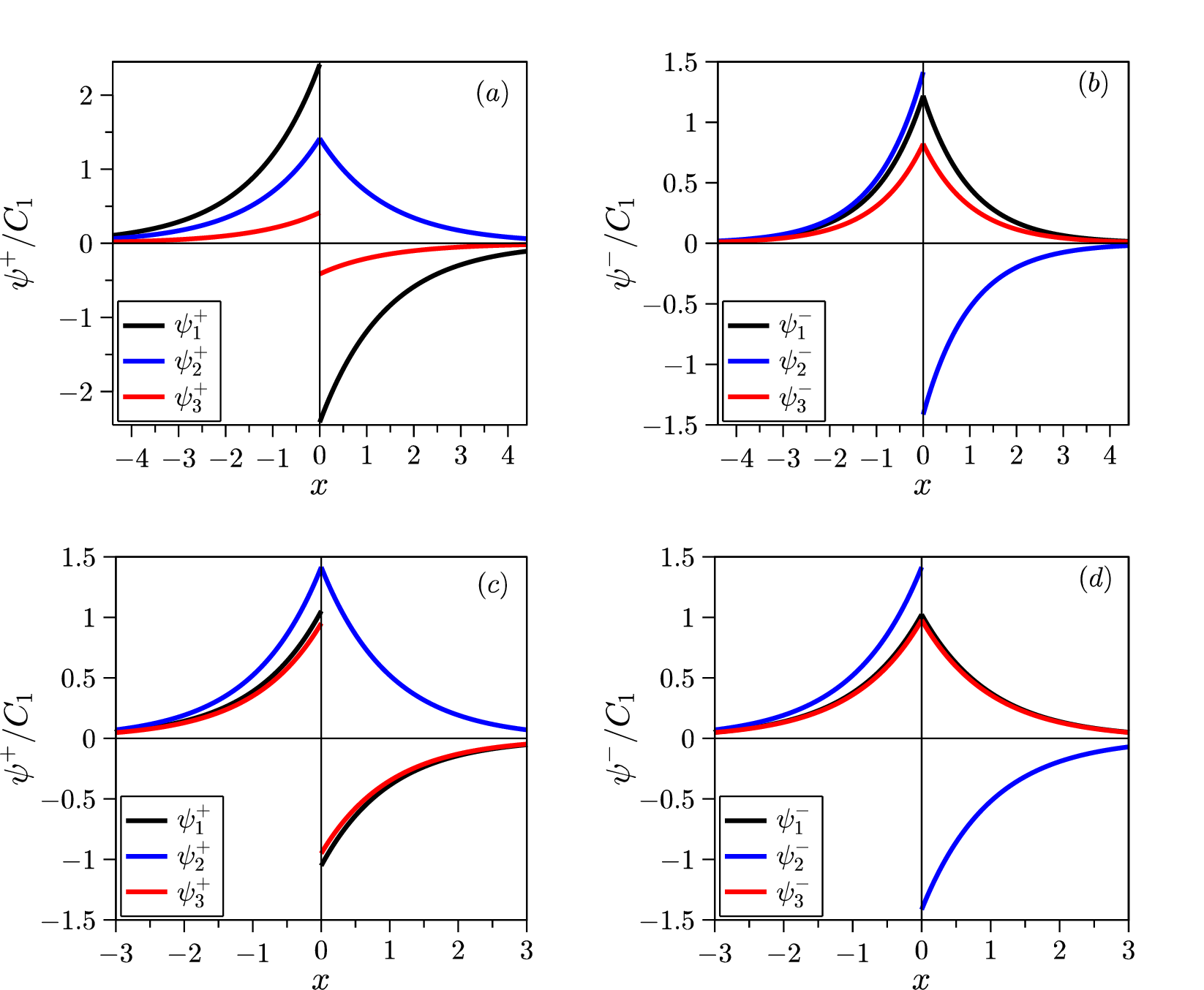}
\caption{Squeezed eigenfunctions $\psi^\pm(x)$ that correspond to
spectrum of type $H$, given by equations (\ref{psi_out_0}) and implemented on set
$A_{H,2}$.  Bound state energies $E_b^\pm =E_n$,
 obtained in the $l^{-2}$-limit,  are  computed
 for the four lowest energy levels:  (a) $n=0$, (b) $n=1$, (c) $n=2$, (d) $n=3$,
in accordance with equations  (\ref{E_H_ground_energy}) for $E_0$ and (\ref{bsII_excited_energies}) for $E_n$, $n=1,\,2,\,3$, with strength $g=2$. }
\label{fig11}
\end{centering}
\end{figure}

Finally, knowing the bound state energies in the squeezing limit
and therefore the values $\rho(E_b^\pm)$, one can plot the eigenfunctions
given by (\ref{psi+-0}). Here, we have restricted ourselves to the bound states
of type $P$ (see figure~\ref{fig10}) and $H$ (see figure~\ref{fig11}).

\section{Concluding remarks}

The energy spectrum of the ordinary  one-dimensional non-relativistic Hamiltonian
for a particle in a constant  potential field $V(x) \equiv V \in \R$
is quite trivial: it is just the shift of a free-particle spectrum by
the strength $V$. The spectrum of the one-dimensional pseudospin-one Hamiltonian
with a constant  three-component potential $V(x) = {\rm col}(V_{11},\,V_{22}, \,V_{33})$
consists of three (upper, middle and lower) bands, which are described by cubic
equation (\ref{law}). The structure of this spectrum crucially depends on
the relative configuration of the strengths $V_{11},\,V_{22}$, $V_{33}$ and all
the possible forms of the bands are illustrated by figure~\ref{fig1}. In this regard,
each strength configuration can be associated with a vector  in
the three-dimensional space $(V_{11},\,V_{22}, \,V_{33})$ and only in the particular case
of the line $V_{11} = V_{22} = V_{33} \equiv V$ in this space,
free-particle spectrum (\ref{free_bands}) with eigenfunctions (\ref{psi_pm_0})
 is shifted by $V$, equally in each band, similarly to the situation with
a non-relativistic Hamiltonian.

Almost all the points in the $(V_{11},\, V_{22}, \, V_{33})$-space correspond to the
energy spectrum, in which the middle band is dispersive. At some limiting points in this space, the middle band as a function of the wave number $k$
shrinks to a line, the so-called flat band, as shown in panels (d) and (h)--(j)
of figure~\ref{fig1}. This set, shown in figure~\ref{fig2}, consists of the two
intersecting planes ${\cal A}$ and ${\cal B}$, which are  defined by
equations (\ref{rel_A}) and (\ref{rel_B}).
 The energy of the upper and lower dispersion bands
on these planes are described by solutions (\ref{E(k)_A}) and (\ref{E(k)_B}).
 The corresponding eigenfunctions are given through the general formula (\ref{gen_sol}).

To implement bound states in the pseudospin-one Hamiltonian,
the components of the potential $V(x)$ must be localized on the $x$-axis.
To this end, we have chosen these components in the form of rectangles
that represent a layer of thickness $l =x_2-x_1$, where $x=x_1$ and $x=x_2$
are arbitrary points. Then one can use the general solution (\ref{gen_sol})
for the interval $x_1 \le x \le x_2 $  complementing it by
the free-particle solution beyond this interval that decreases as $|x| \to \infty$
and using the matching conditions at the edges $x_1$ and $x_2$.
Within this approach, a pair of general equations (\ref{bs_eqs}) has been derived
for finding bound state energies. The solutions to these equations
 are conditionally denoted as $E^+ =E_b^+$ for the first equation (\ref{bs_eqs})
 and $E^- =E_b^-$ for the second one. The  corresponding eigenfunctions
 $\psi^+(x)$ and $\psi^-(x)$ are illustrated by expressions
(\ref{psi+_in})--(\ref{psi-_out}) and figure~\ref{fig3}.

 As demonstrated by figures from \ref{fig4} to \ref{fig9}, the structure
of the bound state spectrum crucially depends on the configuration of the
strengths $V_{11}$, $V_{22}$ and $V_{33}$ that determine the rectangular potentials.
 For simplicity, instead of these three independent strengths as vectors in
 the $\R^3$-space, we have restricted ourselves to the investigation on the two pencils
 of straight lines in $\R^3$ defined by equations (\ref{rep1}) and (\ref{rep2}), where
only  one strength parameter $V$ is incorporated. Even on these particular sets,
a whole variety of bound states has been proven to exist. Based on the
asymptotic behavior of the solutions to equations (\ref{bs_eqs}) in the limit as
$|V| \to \infty$, one can single out the four types of bound states,
which we call in the present work as $P$, $D$, $H$ and $W$.
   The energies for two of these types ($P$ and $H$) have already been investigated
   earlier  in  paper \cite{Zhang2022JPB}. In the present work, the study of these
energies  has been supplemented by  the solutions  with imaginary
wave number $k$. Particularly, for the potentials with all the three strengths
$V_{11}$, $V_{22}$
and $V_{33}$ $\propto$ $V$, the energy spectrum of the type $P$ consists of two levels
and the dependence on $V$ is almost periodic (and exact periodic in the limit
as $|V| \to \infty$). Surprisingly, the energy spectrum of the type $H$ consists of an
infinite number of levels, resembling the hydrogen atom spectrum
($E_{n} \propto 1/n^{2}$, $n=1,2, \ldots$). It should be noticed that
a successive cutoff of energy levels with the growth of the strength $V$
is possible for the type $H$ (compare figures~\ref{fig6} and \ref{fig7}).
 In addition to the types $P$ and $H$,
  we have examined the spectrum that consists of two levels for large $V$
  (type $D$) merging into a single level in the limit as $|V| \to \infty$.
Another behavior of the bound-state energies, which has been observed in the present work,
is a successive detachment
of energy levels from the thresholds of upper and lower continuums $E = \pm m$ with increasing the strength $V$ (type $W$).   This behavior resembles
the energy spectrum of an ordinary potential well as its depth $V$
tends to infinity at fixed width.  The energy levels for this type
 have been shown to behave as  $E_n \propto n^2$, $n=1,2, \ldots$.

For some bound state energy levels, it is possible to use a one-point approximation
through the squeezing limit $l \to 0$  and $V\to\infty$.
To implement such a procedure explicitly,
we consider the strength parameter $V$ as a function of $l$ imposing an appropriate
behavior as $l \to 0$. For realizing a well-defined point interaction, it is sufficient
to construct a matrix that connects the two-sided values of the wave function given
at the point of singularity (for instance, at $x = \pm 0$). To this end,
we have derived the general form of such a matrix for arbitrary points $x_1$ and
$x_2$ given by equation (\ref{L_matrix}). Further, in each special case,
setting $x_1 \to -0$ and $x_2 \to +0$, we have realized the following
three families of one-center point interactions using the asymptotic behavior of $V$
as $l \to 0$: (i) $V \sim g/l$ ($\delta$-limit),
 (ii) $V \sim g(m/l^2)^{1/3}$ ($l^{-2/3}$-limit) and (iii) $V \sim g/l^2m$
 ($l^{-2}$-limit), where $g$ is a dimensionless  coupling constant.

In conclusion, it would be interesting to develop a more general approach for 
studying bound states, which avoids the presentation of the three strengths 
$V_{11}$, $V_{22}$ and
$V_{33}$ through one parameter $V$. In this way, it could be possible to obtain
additional types of bound state spectra. The investigation of the systems with a
long-range Coulomb potential, instead of piecewise potentials, is of big interest 
as well.
The study of the scattering problem in the presence of multiple point potentials,
including resonance effects as well as bound states in the continuum, is also 
of great interest.

\bigskip
 {\bf Data availability statement }

\smallskip
The data that support the findings of this study are available upon
reasonable request from the authors.

\bigskip
 {\bf  Acknowledgments}

\smallskip

We would like to thank the Armed Forces of Ukraine
for providing security to perform this work. A.V.Z. acknowledges financial support from the National Academy of Sciences of Ukraine, Project No.~0123U102283.
Y.Z. and V.P.G. acknowledge financial support by the National Research
Foundation of Ukraine grant (2020.02/0051) `Topological phases of matter and excitations in Dirac materials, Josephson junctions and magnets'.
Finally, we are indebted to the anonymous Referees
for the careful reading of this paper, their questions and suggestions, resulting
in the significant improvement of the paper.

 \end{document}